\documentclass[aps,prx,superscriptaddress,floatfix,showpacs,notitlepage,twocolumn,titleinbib]{revtex4-2}

\usepackage{graphicx}% http://ctan.org/pkg/graphicx
\usepackage{ragged2e}

\usepackage{mdframed}
\newmdtheoremenv{theo}{Theorem}

\usepackage[labelfont=bf]{caption}
\usepackage{graphicx}% Include figure files
\usepackage{dcolumn}% Align table columns on decimal point
\usepackage{subfig}
\usepackage{bm}% bold math
\usepackage{amsmath}
\usepackage{hyperref}% add hypertext capabilities
\begin{document}

\preprint{APS/123-QED}

\title{Adiabatic state preparation and thermalization of simulated phase noise in a Rydberg spin Hamiltonian}% Force line breaks with \\
%\thanks{A thanks to all the collegues that made this work possible with their stimulating conversation}%
\author{Tomas Kozlej}
\affiliation{University of Strathclyde, 16 Richmond St, Glasgow G1 1XQ%
}%
% \email{kozlej.tomas@strath.ac.uk}
\author{Gerard Pelegri}
\affiliation{University of Strathclyde, 16 Richmond St, Glasgow G1 1XQ%
}%
\author{Jonathan D. Pritchard}
\affiliation{University of Strathclyde, 16 Richmond St, Glasgow G1 1XQ%
}%
\author{Andrew J. Daley}

\affiliation{University of Strathclyde, 16 Richmond St, Glasgow G1 1XQ%
}%
\affiliation{%
University of Oxford, Wellington Square, Oxford OX1 2JD
}%

\date{\today}
\begin{abstract}
Laser phase noise is one of the main sources of decoherence in driven Rydberg systems with neutral atoms in tweezer arrays. While the effect of phase noise in the regimes of isolated qubits and few-qubit gate protocols has been studied extensively, there are open questions about the effects of this noise on many-body systems. In many scenarios, the effects of noise cannot simply be described by an increase in the energy or temperature of the system, leading to non-trivial changes in the state and relevant correlations. In this work, we use stochastic sampling to simulate laser phase noise based on experimentally relevant spectral densities. We explore the impact of this noise on adiabatic state preparation in a one-dimensional system, discussing the interplay between dephasing and diabatic excitation during dynamics. Moreover, we investigate the approximate thermalization of dephasing energy in an interaction regime resonant with the noise, showing convergence to canonical ensemble predictions for two relevant observables.
\end{abstract}

\maketitle

%\tableofcontents

\section{Introduction}

Progress on the scalability of optical trapping and coherent control have made neutral atom arrays an exciting platform for quantum computation and simulation \cite{Saffman2010,Barredo2016,Henriet2020,Morgado2020}. In these systems, strong inter-particle interactions between individually trapped atoms are realized by exciting atoms to high-energy Rydberg levels, giving rise to a blockade effect \cite{Johnson2008} that can be leveraged for the creation of quantum gates \cite{Lukin2001,Ebadi2020,Pelegrí_2022}, as well as to a range of spin Hamiltonians \cite{Bernien_2017,Ebadi2022}. Applications of these systems include the exploration of annealing processes relevant for optimization \cite{Serret_2020,Nguyen_2023}, realizing topological models \cite{de_Leseleuc_2019,Weber2017}, spin liquid phases \cite{Semeghini_2021,Giudici_2023,Kornja_a_2023}, quantum many-body scars \cite{Bluvstein_2022} and information scrambling \cite{Hashizume_2021,scrambling_2024,Bluvstein_2023}. Much of the recent progress in this area came from reductions in laser phase noise, which crucially affects excitations to Rydberg levels and directly generates dephasing \cite{Leseleuc18,Levine2018,day2021,Jiang_2023}. The effects of this laser phase noise have mainly been characterized in the context of single qubit operations. However, in simulation of quantum dynamics, an important question this paper asks is how this noise affects the overall behavior of a many-body system. This connects to fundamental questions about thermalization in closed quantum systems \cite{Rigol_2009,Cassidy_2011,Schachenmayer_2015,D_Alessio_2015,Deutsch_2018}, specifically whether heating due to laser phase noise results in correlation functions predicted by the increase in energy and temperature.

In this work, we quantify the effects of laser phase noise on many-body atomic spin systems as realized in tweezer arrays. The adiabatic preparation of many-body ground states provides  a simple but relevant test bed for our study of laser phase noise, used both to prepare interesting many-body states \cite{Pohl_2010_CrystallineExperiment,Schach2010,Schaub_2015,Bernien_2017}, as well as in quantum annealing \cite{Kadowaki_1998,Ebadi2022,Glaetzle_2017,Kim2024}. We simulate the time evolution of an adiabatic state preparation of our antiferromagnetic spin chain, applying laser phase noise that has been sampled from experimentally realistic diode laser data, similar to phase noise measured, e.g., in Ref. \cite{Leseleuc18}. Furthermore, to avoid complicating the discussion and obscuring the effects of noise, we work with a linear adiabatic ramp in the dipole interaction regime, typical of early theoretical \cite{Schach2010} and experimental demonstrations \cite{Pohl_2010_CrystallineExperiment} of this adiabatic protocol. We extract features that are relevant to general noise spectra, using the loss of ground state fidelity to quantify the competition between diabatic and dephasing excitation that occurs when we change laser parameter variation rates, as well as showing the effect of changing noise strength, and interaction strength between sites. We supplement these results with a nuanced look into excitation dynamics in our noise, providing an analysis on available transition elements that occur at different stages of the protocol, and how these lead to distinct excitation patterns in time independent Hamiltonian evolutions. Next, we ask to what extent can the energy added to the system by phase noise lead to apparent thermalization in two relevant observables. To answer this question, we work in an interaction regime that is considerably more resonant with the applied noise, allowing us to clearly observe mechanisms through which dephasing excites a many-body system. We evaluate the long-time average expectation values of relevant observables, comparing them to predictions by a thermal equilibrium state of the same energy to see how closely the relaxation of energy resembles thermalization.

The manuscript is organized as follows: after an introduction to the physical system and model in Section \ref{Sec 2}, Section \ref{Sec 3} considers the effects of phase noise on the simulations of an adiabatic state preparation of a one-dimensional Rydberg spin chain with antiferromagnetic ($Z_2$) ordering. Section \ref{Sec 4} introduces the concept of thermalization, followed by a demonstration of thermalization in two separate many body observables. Finally, Section \ref{Conclusion} provides a concluding discussion on the analysis of laser phase noise in Rydberg systems conducted in this paper. We contextualize the findings within the broader advances in state-of-the-art neutral atom systems, highlighting how insights gained from this work underscore the importance of understanding and mitigating phase noise effects in many-body quantum systems.

\begin{figure}[t!]
\hspace{-15pt}\includegraphics[width=8.6cm]{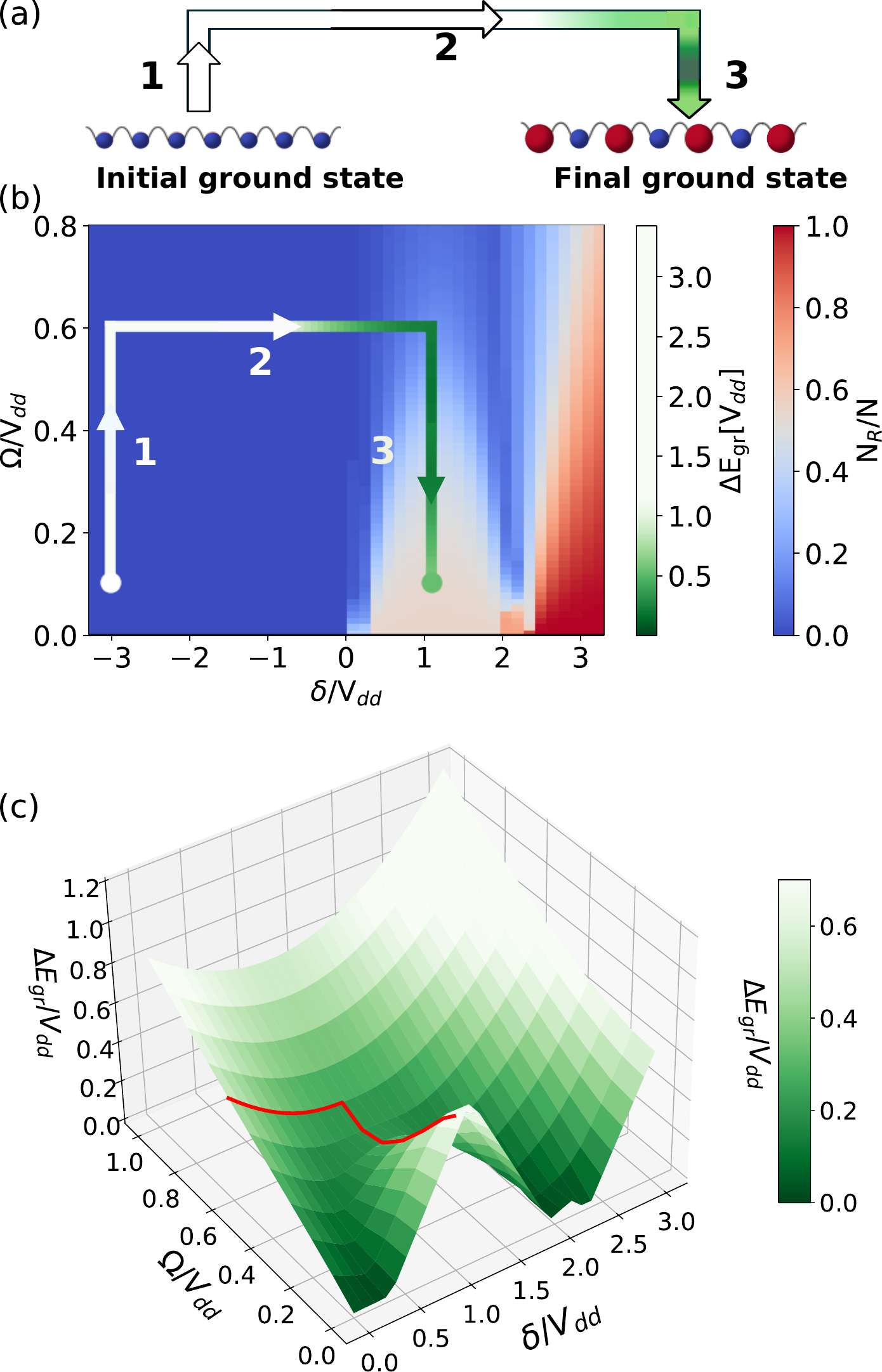}
\caption{\label{fig:Ryd Adiab Sketch} Laser parameter space plot and ground state energy gap surface plot for the Rydberg Hamiltonian seen in Eq. \eqref{Ryd Hamil} for a chain of $N=11$ atoms. (a) A schematic diagram showing crystalline ordering of ground state (blue) and Rydberg excited (red) atoms for initial and target many-body state of adiabatic state preparation. (b) Total fraction of excited Rydberg atoms $N_R/N$ in the ground state plotted against laser parameters $\Omega$ and$\delta$, from no excitations (blue), to half filled (white), to fully excited (red).\justifying The path for the diabatic state preparation studied is provided with color gradient from white to green tracking ground state energy gap. (c) Surface plot of the ground state energy gap in the region of positive $\delta/V_{dd}$, with the path of adiabatic state preparation in this region highlighted in red.}
\end{figure}

\section{The physical system \label{Sec 2}}

The simulations discussed in this paper are carried out in a one-dimensional Rydberg spin Hamiltonian of the form

\begin{flalign}
\label{Ryd Hamil}
\hat{H}_{R} =\quad& \frac{\Omega(t)}{2}\sum^N_k\big[e^{-i\phi(t)}|0\rangle_k \langle 1|_k+e^{i\phi(t)}|1\rangle_k \langle 0|_k ]\nonumber \\ &- \delta(t) \sum^N_k \hat{n}_k + V_{dd} \sum ^N_{k,l> k}\frac{\hat{n}_k\hat{n}_l}{|k-l|^3},
\end{flalign}

where the first term drives excitations between the ground $|0\rangle$ and Rydberg $|1\rangle$ states on site $k$ along the chain, at a Rabi frequency $\Omega(t)$ modulated by a phase noise signal $\phi(t)$ that introduces a small error at every time step. The parameter $\delta(t)$ models the global laser detuning, and the operator $n_k\equiv\frac{1}{2}(\sigma_z^k-\hat{I})$ counts the number of Rydberg excitations in site $k$. The terms $\delta(t)$ and $\Omega(t)$ are tuneable in an experimental setting and are governed by the laser frequency, amplitude, and phase stability. Finally, we define the interaction Hamiltonian $\hat{H}_{\text{int}}$ explicitly to use it in Section \ref{Sec 4} as an observable

\begin{eqnarray}
\label{H_int}
\hat{H}_{\text{int}}=  V_{dd} \sum ^N_{k,l> k}\frac{\hat{n}_k\hat{n}_l}{|k-l|^3},
\end{eqnarray}

introducing a dipole-dipole interaction strength $V_{dd}$ mediated by $\hat{n}_k\hat{n}_l$ interactions that enforce a dipolar blockade penalty on neighboring excitations. We work in the dipole dipole interaction regime in which Rydberg states must maintain a permanent dipole moment that is typically achieved through strong electric fields \cite{Ates_2008,Browaeys_2016,Leseleuc17}, and long range interactions decay as a function of distance between sites $d^3=|k-l|^3$, where $k$ and $l$ are site indices and the lattice constant is absorbed into $V_{dd}$. Open boundary conditions are considered for the one-dimensional chain, for which atoms found on the edges of the chain experience less of the effect of Rydberg blockade and are therefore more likely to be Rydberg excited than atoms within the chain. This means that energetically favorable $Z_2$ ordering of excitations can be identified uniquely in chains with an odd number of lattice sites, but an even number of lattice sites results in a central domain wall after which ordering flips \cite{Bernien_2017}. To simplify notation and discussion, here we will focus on odd numbers of spins. Furthermore, while the noiseless protocol can be generalized across different frequency regimes, for computational simplicity we define an experimentally reasonable relative number to which we set the interaction strength $V_{dd}/2\pi=10$~MHz \cite{Ravets2014,Bernien_2017}, and use this to scale all units in the system with $\hbar =1$ also assumed. Hence, both $\Omega$ and $\delta$ are expressed in units of this relative $V_{dd}$, while time is given in units of $V_{dd}^{-1}$. Note that for $\Omega=0$ the Hamiltonian in Eq. \eqref{Ryd Hamil} is reminiscent of the Ising model for magnetic dipoles. In such a regime, the competition between $\delta$ and $V_{dd}$ leads to an excitation ladder in which ground states with increasing number of evenly spaced excitations are favored \cite{Pohl_2010_CrystallineExperiment,Schaub_2015,Petrosyan2017}. These crystalline ground states exhibit large energy gaps that make them robust to small increases in $\Omega$, leading to regions in the $\Omega/\delta$ parameter space where the ground state is dominated by a single crystalline state. Figure \ref{fig:Ryd Adiab Sketch}(a) shows a diagram of the ground state of the initial and final Hamiltonian for the three part adiabatic state preparation discussed in Section \ref{Sec 3}, while in Figure \ref{fig:Ryd Adiab Sketch}(b) we show the path that this protocol takes along the $\Omega$ and $\delta$ parameter plot. The figure also shows Rydberg excitations along the spin chain as a function of $\Omega$ and $\delta$ parameters, clearly displaying the formation of a lobe where a stable ground state with crystalline $Z_2$ order of excitations persists. How close the final prepared many-body ground state $|\psi_{gr}\rangle$ is to a perfect $Z_2$ ordering can be tested by computing the expectation value of the order parameter 

\begin{equation}
\label{Z2}
    \hat{O}_{Z_2}=\sum^N_{k,l\neq k}  \sigma_z^k (-1)^{k+l}\sigma_z^l,
\end{equation}

\noindent where we check anti-ferromagnetism over all different lattice sites $k$ and $l$. The goal of the procedure is thus to take a ground state with no Rydberg excitations to a ground state with $Z_2$ ordered excitations and expectation value $\langle \hat{O}_{Z_2}\rangle=1$, while maximizing the ground state energy gap to mitigate diabatic excitation. Finally in Figure \ref{fig:Ryd Adiab Sketch}(c) we also provide a surface plot of the ground state energy gap $\Delta E_{gr}/V_{dd}$ again as a function of $\Omega/V_{dd}$ and $\delta/V_{dd}>0$, along with a highlighted path taken through this region by the adiabatic protocol. The surface plot shows the emergence of a saddle point as $\Delta E_{gr}/V_{dd}$ drops at the transition point to a $Z_2$ ordered ground state, resulting in a critical region where low $\Delta E_{gr}/V_{dd}$ could result in increased excitation during a linear adiabatic ramp. The next section will introduce this state preparation in more detail as well as providing results after introducing phase noise to the system.

\section{Phase noise in adiabatic state preparation\label{Sec 3}}

\begin{figure}[t]
\includegraphics[width=8.6cm]{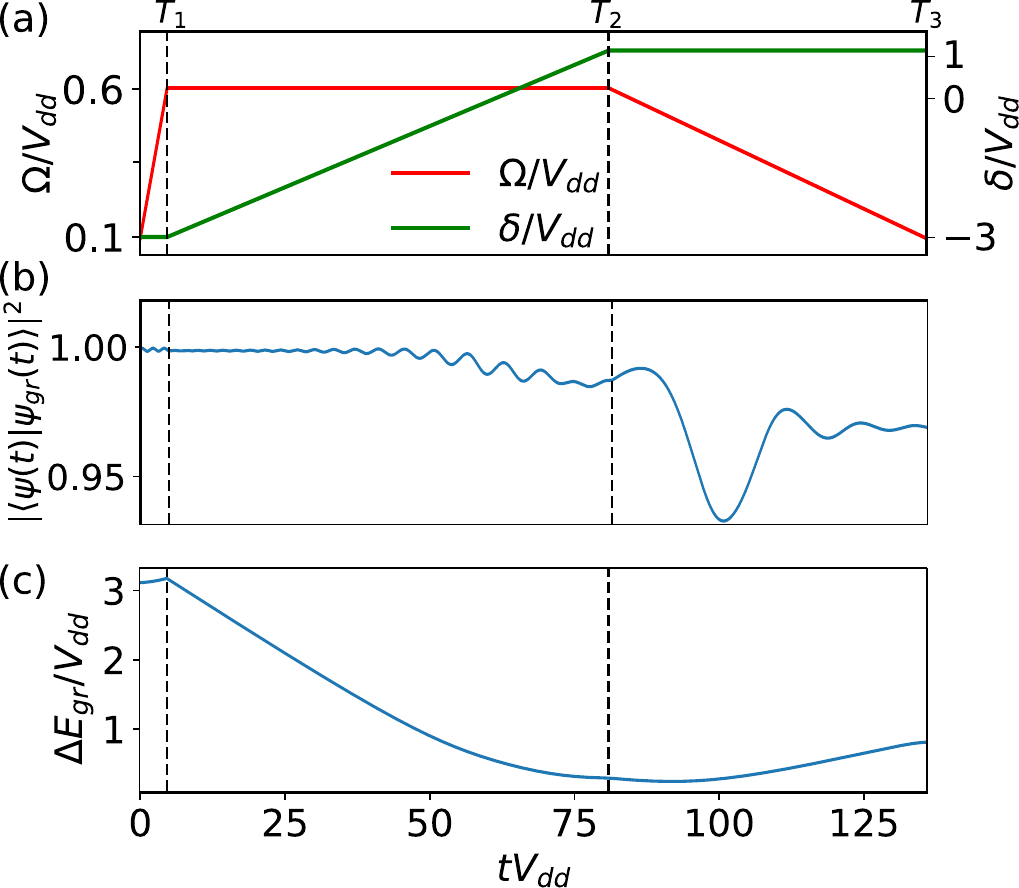}
\caption{\label{fig:RAMP123} (a) Laser parameters for three step adiabatic state preparation of a $Z_2$ ordered crystalline state for a 11 atom Rydberg chain. Rabi frequency $\Omega$ (red) is tuned linearly in step 1,3 for periods $T_1$, $T_3$, while detuning $\delta$ (green) is tuned linearly only in step 2 for a period of $T_2$. (b) Concurrent fidelity measured as the wavefunction overlap between the ground state $|\psi_{gr}\rangle$ of the instantaneous Hamiltonian $\hat{H}_{R}(t)$ and the evolved state $|\psi\rangle$. \justifying (c) Energy gap $\Delta E_{gr}$  between the instantaneous ground state and the first excited state.  }
\end{figure}
Adiabatic state preparation is a technique to initialize a quantum system in a desired state \cite{Petrosyan2017,Schach2010}. The method is useful for preparing complex quantum states that are otherwise difficult to generate directly. It typically involves slowly varying the controllable parameters of a system to transform a given energy eigenstate state in an initial Hamiltonian to a target state that is the ground state of the modified Hamiltonian. This variation should be slow enough to satisfy the adiabatic constraint 

\begin{eqnarray}
|\langle m(t)|\dot{H}(t)|n(t) \rangle| \ll |E_m(t)-E_n(t)|^2,\\
\nonumber
\end{eqnarray}

\noindent  where we maintain $\hbar=0$ and with $|m(t)\rangle$ and $|n(t)\rangle$ being eigenstates corresponding with energies $E_m(t)$ and $E_n(t)$ in the instantaneous Hamiltonian with an energy gap that we do not wish to cross. Typically, the procedure is done starting from a state which is easy to prepare experimentally to a target state which is harder to access. The challenge in adiabatic state preparation is thus to maximize the energy gap to the first excited state $\Delta E_{gr}$ and to decrease the variation rate if $\Delta E_{gr}$ becomes small enough to result in unwanted excitation of higher energy eigenstates.

\subsection{The adiabatic ramp}

\begin{figure}[b]
\includegraphics[width=8.6cm]{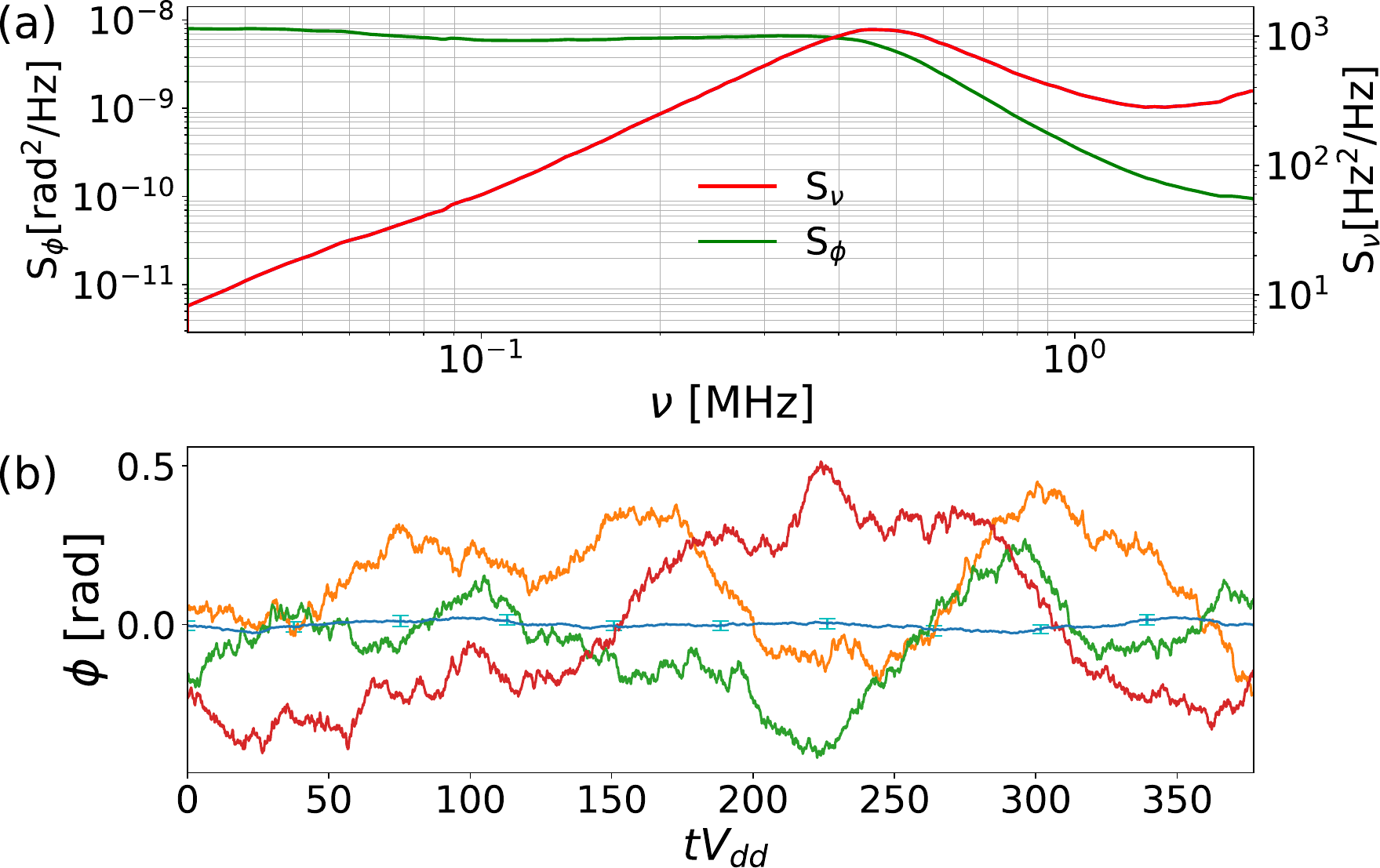}
\caption{\label{fig:My PSD} Realistic noise power spectrum used to generate independent phase noise time realizations implemented in adiabatic time evolution. (a) The phase noise spectrum $S_\phi$ (green) which has been converted from a frequency spectrum $S_\nu$ (red) used in noise generation. The relevant conversion equation is $S_\phi=\nu^2S_\nu$. (b) Three independent noise realizations generated using $S_\phi$ (see Appendix \ref{NoiseGEN}) provided in units of $tV_{dd}$ where $V_{dd}/2\pi=10$MHz. \justifying The averaged noise behavior after 100 independently generated noise realizations is shown in blue. }
\end{figure}

\begin{figure*}[t]
    \centering
    \includegraphics[width=0.95\textwidth]{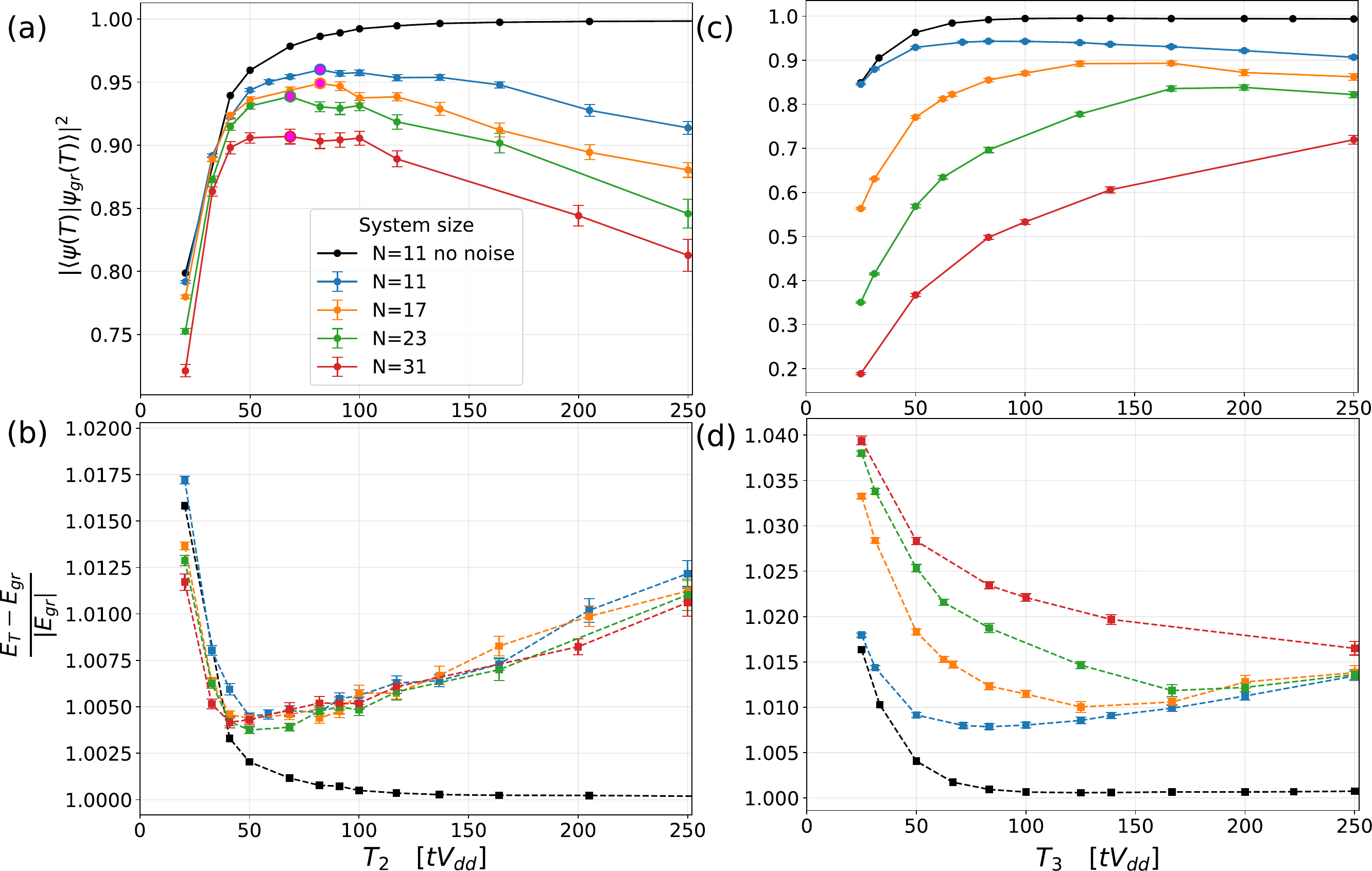}
    \caption{\label{fig:noisy Rtime}
    Final fidelity of prepared state $|\psi(T)\rangle$ with the instantaneous ground state $|\psi_{gr}(T)\rangle$ at the end of the ramp, and final energy $E_T$ relative to the ground state energy $E_{gr}=\langle\psi_{gr}| H_R|\psi_{gr}\rangle$ after simulations using different evolution times $T_2$ for stage 2 (a,b), and $T_3$ for stage 3 (c,d) in the adiabatic state preparation described in Figure \ref{fig:Ryd Adiab Sketch} with added laser phase noise. \justifying Colored lines show results for noisy state preparation with a variety of system sizes, while the black line provides a reference for 11 sites and no added noise. Optimal values of $T_2$ resulting in highest fidelity for a given $N$ are highlighted in larger magenta circles in (a), with their final states used for $T_3$ analysis in (c,d). Each data point has error bars showing associated standard error after averaging over 100 preparations with unique laser phase noise signals generated from the same power spectrum in Figure \ref{fig:My PSD}. These results have been calculated using tensor network simulations using the iTensor package in Julia \cite{itensor}, with time evolution achieved using the TDVP algorithm with a maximum bond dimension of 100.}
\end{figure*}

Both laser parameters $\delta$ and $\Omega$ in the Hamiltonian in Eq. \eqref{Ryd Hamil} are time dependent and can be tuned independently. Starting from a Hamiltonian with large negative detuning in a ground state with no Rydberg excitations, Figure \ref{fig:RAMP123}(a) shows the preparation of the $Z_2$ target state using a variation of the three-step adiabatic ramp originally proposed in \cite{Schach2010}. The initial Hamiltonian is far detuned at laser parameters $\Omega/V_{dd}=0.1$ and $\delta/V_{dd}=-3$ where a large ground state gap allows for reliable experimental preparation. The target Hamiltonian occurs at $\Omega/V_{dd}=0.1$ and $\delta/V_{dd}=1.1$ with a much smaller gap and a ground state with $Z_2$ excitation ordering. Figure \ref{fig:RAMP123}(a) plots the linear tuning of the two laser parameters across all three stages of the protocol, while Figures \ref{fig:RAMP123}(b,c) show the change in ground state fidelity $|\langle\psi|\psi_{gr}\rangle|^2$ and ground state energy gap $\Delta E_{gr}$  of such an adiabatic state preparation for a chain of 11 atoms. Since the evolution time is finite and thus not perfectly adiabatic, there is a small amount of excitation throughout the process leading to a drop in overall fidelity. At the transition between stages 2 and 3 $\Delta E_{gr}$ decreases considerably as shown in Figure \ref{fig:Ryd Adiab Sketch}(c), dramatically increasing the likelihood of exciting higher many-body energy states. Most unwanted excitations occur during this critical region in the adiabatic preparation, causing a notable fidelity drop that can only be mitigated by slowing the parameter variation rate. The following section will therefore test this state preparation for a variety of variation rates across stages 2 and 3, while also exposing them to phase noise.

\subsection{Introducing phase noise}

When noise is added to the transverse field, the resulting dephasing of the dynamics accumulates and, over time, causes unwanted excitations. Noise is added to the system with $\phi(t)$, which is implemented as a phase term modulating $\Omega(t)$ in Eq. \eqref{Ryd Hamil}. Realizations of $\phi(t)$ are generated using an adaptation of the TK95 algorithm \cite{TK95} that stochastically samples a power spectrum chosen to be typical for Rydberg lasers. Figure \ref{fig:My PSD}(a) provides the original frequency power spectrum of laser noise for a diode laser, and the corresponding phase power spectrum that can then be used to generate the unique discrete noise signals shown in Figure \ref{fig:My PSD}(b). Details on the simulation of realistic phase noise and the excitation dynamics that lead to the results in this paper are provided in the Appendices \ref{NoiseGEN} and \ref{noiseEXC}. 

Figure \ref{fig:noisy Rtime} shows how different adiabatic evolution times result in a drop in fidelity of the final prepared state with the instantaneous ground state at the end of stage 2 in plot (a), and at the end of stage 3 in plot (c) which marks the end of the adiabatic protocol. Note that the evolution time of stage 1 is fixed to $T_1V_{dd}=5$ as this is sufficient to preserve high fidelity. To generate data for stage 3 we use the prepared state at the end of stage 2 that corresponds to the highest fidelity. In addition, the energy of the final prepared state relative to the ground sate is also provided, in (b) for stages 1 and 2, and in (d) for stage 3. Each data point in Figure \ref{fig:noisy Rtime} thus represents a final fidelity or energy at the end of adiabatic ramps simulated for a separate evolution time and system size, averaged across 100 independent noise realizations of phase noise, while a black line corresponds to results for an 11-site system without laser phase noise for reference. When no laser noise is present, we see continued improvement in fidelity as the evolution time increases and diabatic excitations are suppressed. This improvement is rapid for short evolution times and more gradual in the case of longer ramp times. However, the addition of phase noise causes an overall drop in fidelity for preparation across all evolution times, and in particular for longer evolution times where the cumulative dephasing leads to prolonged excitation out of the ground state. This difference in fidelity and energy is present for both stages and is best observed for the system size of $N=11$, with a clear divergence between realizations with and without noise.

Looking at the final fidelity of the prepared state at the end of stage 2, Figure \ref{fig:noisy Rtime}(a) shows the competition between diabatic and dephasing excitations, since both have very different excitation profiles and opposite relationships to evolution time. While diabatic excitation involves low energy eigenstates and occurs for short evolution times after loss of adiabaticity during fast $\Omega$ and $\delta$ variation rates, dephasing excitation can affect higher energy states and occurs after the gradual accumulation of phase errors in the laser drive affecting longer evolution times. This gradual accumulation is precisely why the standard error in Figure \ref{fig:noisy Rtime} increases with time and individual trajectories begin to diverge as independent phase errors build up. The competition between the two excitation mechanisms leads to the emergence of an optimal time $T_{2O}$ (marked by magenta dots in Figure \ref{fig:noisy Rtime}(a)), which is long enough to suppress most diabatic excitation but short enough to allow only small dephasing, thus corresponding to the largest achievable fidelity for a given system size. For stage 2 of the adiabatic state preparation $T_{2O}$ remains roughly consistent between different system sizes, as diabatic excitation remains low except for the shortest evolution times, and the frequency profile of the added noise is the same across all simulations, meaning that the generated noise signals take approximately the same time to begin affecting dynamics. Figure \ref{fig:noisy Rtime}(b) shows the transition from mainly diabatic to mainly dephasing excitation in stage 2, which occurs at a comparatively short evolution time of $T_2V_{dd}\approx60$ for all system sizes, as the ground state energy gap during this ramp remains large enough to permit faster variation rates without inducing excessive diabatic excitation. For stage 2 ramps that are longer than $T_2V_{dd}\approx60$ diabatic excitation becomes insignificant, and we see a steady accumulation of energy from phase noise that is similar across all system sizes since the same noise profile is being used. Even though the same phase noise energy is added into the system, fidelities drop more for larger system sizes which have a larger density of states and smaller energy gaps leading to more excitation out of the ground state. In the case of energy, we see from Figures \ref{fig:noisy Rtime}(b) that the differences in the amount of energy added to the system between different system sizes are mainly governed by diabatic excitation. Since diabatic excitation across the first two stages is minimal due to a large ground state energy gap, differences between energies only occur for very short evolution times where diabatic excitation is the larger for larger systems since they have a higher density of states and smaller energy gaps overall. For longer adiabatic ramps we see energy rise linearly as a function of the time that phase noise has to accumulate. Moving on to stage 3 in Figures \ref{fig:noisy Rtime}(c,d), we see that the dynamics become very different due to the dramatic increase in diabatic excitation as the system passes through a critical region in the parameter space where the ground state energy gap becomes very small. In Figure \ref{fig:noisy Rtime}(c) this increase in diabatic excitation leads to much larger losses in fidelity that are more acute for larger system sizes, as they exhibit a larger density of states and smaller energy gaps. The improvements in fidelity from mitigating diabatic excitation when moving from short to longer evolution times are also much more gradual in stage 3 than across stages 1 and 2 where ramps shorter than $T_2V_{dd}=50$ were strongly affected and the optimal ramp time (which signals a transition from diabatic to dephasing excitation) remained consistently around $T_2V_{dd}\approx75$. Instead, in stage 3 we observe that the optimal evolution times increase with system size, from $T_3V_{dd}=100$ for 11 sites, to $T_3V_{dd}=167$ for 17 sites, and $T_3V_{dd}=200$ for 23 sites, with the 31 site system yet to reach its zenith after a longest simulated time of $T_3V_{dd}=250$. Larger systems sizes are increasingly more susceptible to diabatic excitations and the variation rate must reduce considerably even at the cost of more dephasing. This is consistent with recent analysis \cite{Petrosyan_2016}, predicting exceedingly long adiabatic ramps for larger system sizes due to diabatic losses.

\begin{figure}
\includegraphics[width=8.2cm]{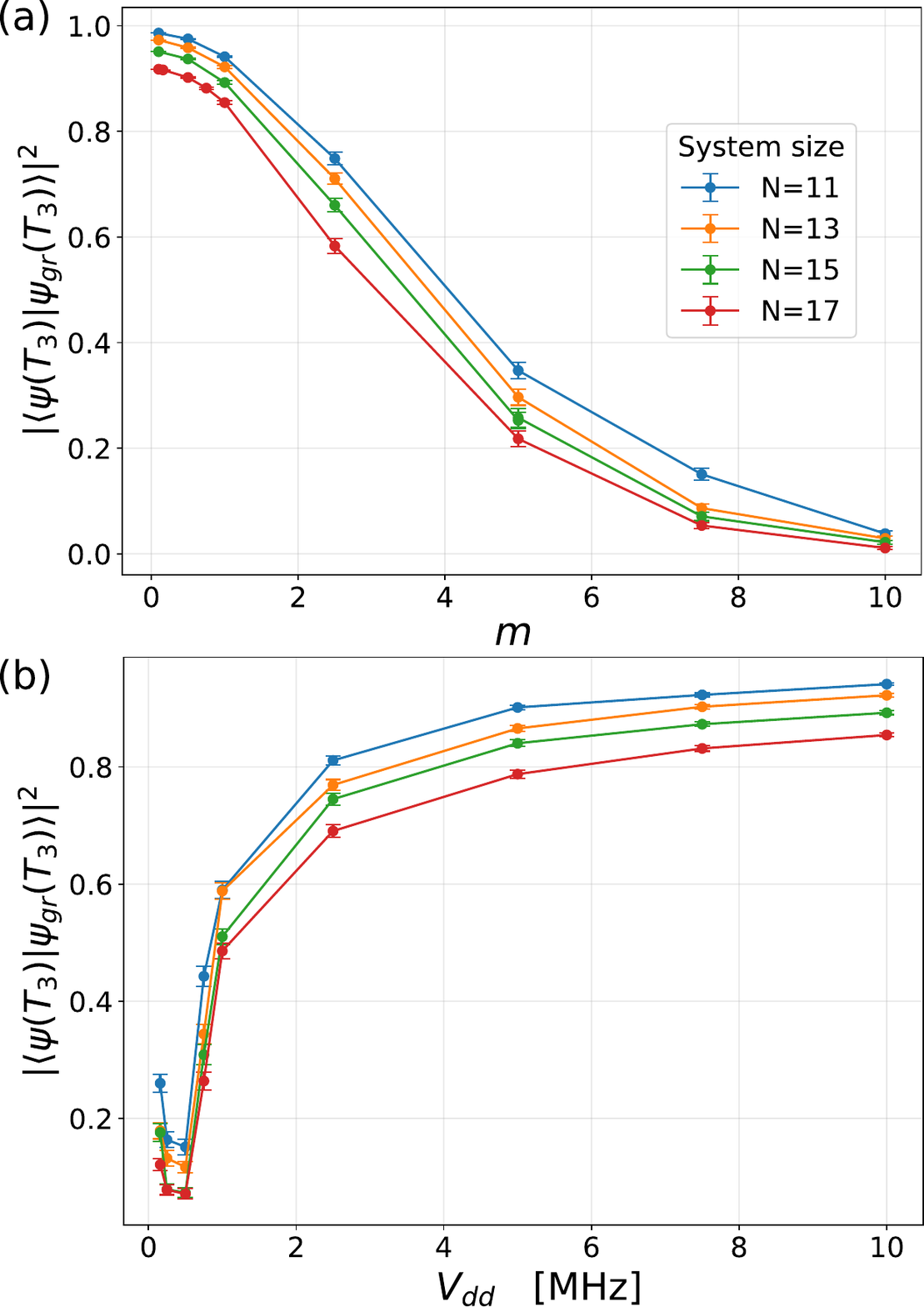}
 \caption{\label{fig:NOISEPoW and Vdd shift}  
 Final fidelities of the prepared state $|\psi(T_3)\rangle$ with $Z_2$ ordered ground state $|\psi_{gr}(T_3)\rangle$ for the adiabatic state preparation described in Figure \ref{fig:Ryd Adiab Sketch} with optimal evolution times for stage 2 ($T_2V_{dd}=82$) and stage 3 ($T_3V_{dd}=83$), and two varying parameters. (a) Final fidelity for different linear scalings of magnitude $m$ are applied directly to phase noise $\phi(t)\rightarrow m\phi(t)$. (b) Final fidelities for different values of $V_{dd}/2\pi$ ranging from $0.1-10$MHz. Colored lines show results for different system sizes. \justifying Error bars show standard error after averaging 100 preparations with unique laser phase noise signals. These results have been calculated using tensor network simulations using the iTensors package in Julia \cite{itensor}, with time evolution achieved using the TDVP algorithm with a maximum bond dimension of 100.
 }%
\end{figure}

Continuing the analysis, Figure \ref{fig:NOISEPoW and Vdd shift} demonstrates the impact that changes in $V_{dd}$ and noise strength have on the performance of the adiabatic protocol for different system sizes while running with optimal variation rates.  To begin with, Figure \ref{fig:NOISEPoW and Vdd shift}(a) demonstrates how the protocol performs for different noise strengths by introducing a factor $m$ that linearly scales the phase noise to $\phi(t)\rightarrow m\phi(t)$. The susceptibility of such adiabatic protocols to phase noise is clear, as we observe an almost complete degradation in fidelity with the target ground state within a single order of magnitude of amplification. Figure \ref{fig:NOISEPoW and Vdd shift}(b) shows the final fidelities after performing the protocol for a variety of $V_{dd}$ values which, given that the laser parameters used are expressed in units of $V_{dd}$, shift the energy scaling of the entire system. Since the frequency profile of the noise remains constant with a broad peak in power at $480$kHz (see Figure \ref{fig:My PSD}), tuning $V_{dd}$ directly shifts the transitions to higher energy levels in and out of resonance. Phase noise made up of frequencies that are too high or too low relative to the dynamics of the system inhibits the transfer of power into the system, thereby causing less excitation out of the ground state.  This is why Figure \ref{fig:NOISEPoW and Vdd shift}(b) shows that the lowest energy added to the system, and thus the highest fidelity achieved, occurs as far as possible from the peak noise frequency at $V_{dd}=10$MHz, with a dramatic degradation in fidelity as the system dynamics move towards affected frequencies at $V_{dd}=0.5$MHz, and a visible improvement as they are surpassed. The result motivates the choice of $10$MHz as a reasonable regime to test the performance of the protocol in Figure \ref{fig:noisy Rtime}, and demonstrates that even if phase noise is present in a laser, the effects of phase noise on many-body experiments can be effectively mitigated by moving dynamics far away from affected frequencies. A similar conclusion was previously shown for single atoms and two qubit gates in \cite{Leseleuc18,Levine2018}.

Studying the effects of phase noise on fidelity provides useful experiment-specific insight on imperfect adiabatic ramps in many-body systems. However, the measure considers only the ground state, revealing little about how dephasing distributes excitations across the energy eigenstates of the system. To get a deeper understanding of the mechanisms through which dephasing excitation propagates we have to analyze the full Hilbert space. The following section provides an analysis of the excitation dynamics for laser phase noise by studying the full energy spectrum of a one-dimensional chain of Rydberg atoms.

\subsection{Matrix element analysis}

\begin{figure}
  \includegraphics[scale=0.4]{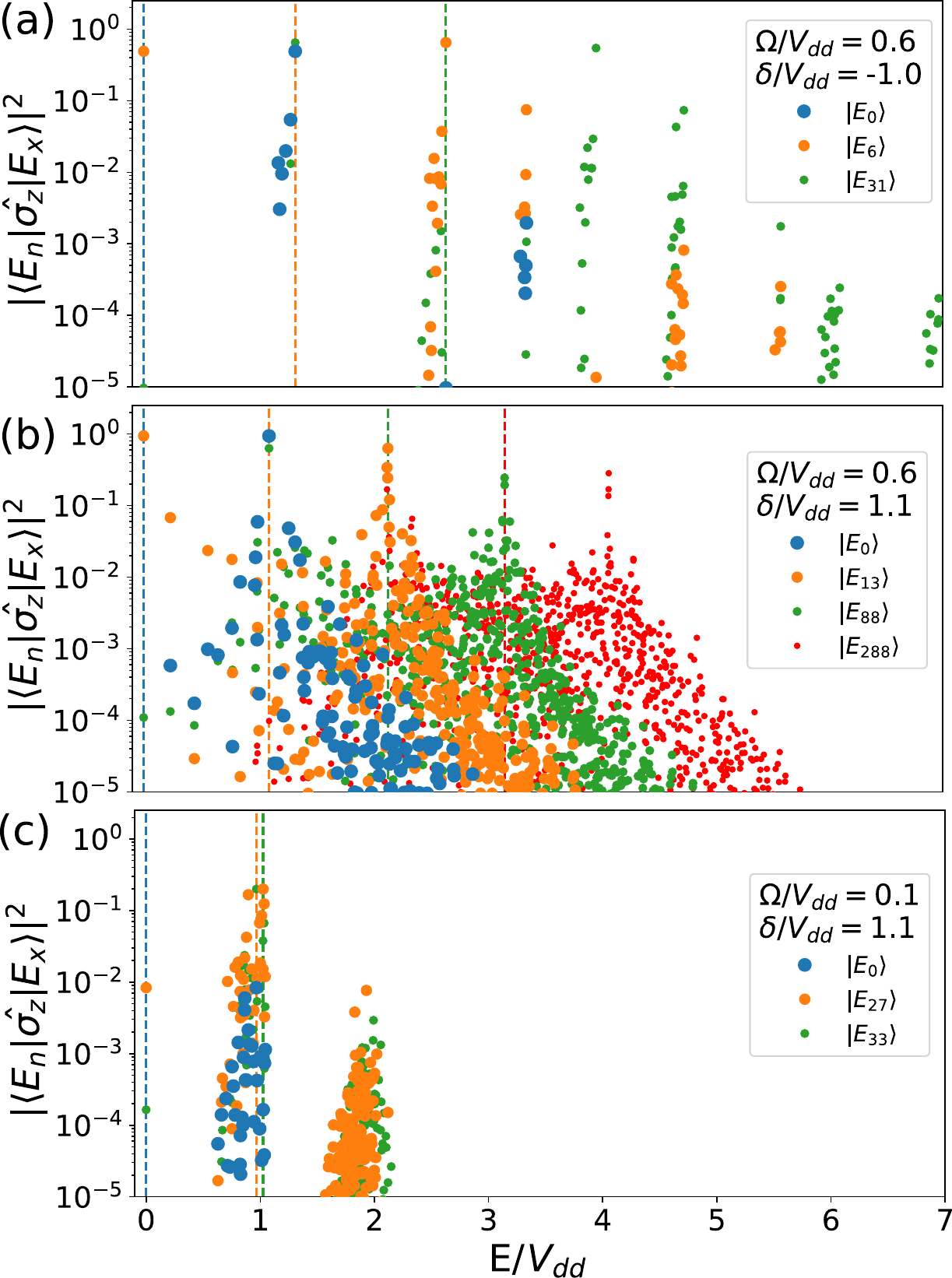}
  \caption{ \label{fig: Matrix Elements} Matrix elements of the many-body operator $\hat{\sigma}_z=\sum_k ^N\hat{\sigma}^k_z$ for Hamiltonians occurring at different stages of the adiabatic state preparation from Figure \ref{fig:Ryd Adiab Sketch} for an 11 site spin chain. Hamiltonian (a), with laser parameters $\Omega/\text{V}_{dd}=0.6$ and $\delta/\text{V}_{dd}=-1.0$ occurs in the middle of stage 2 of the protocol. Hamiltonian (b), with parameters $\Omega/\text{V}_{dd}=0.6$ and $\delta/\text{V}_{dd}=1.1$, occuring at the transition point between stages 2 and 3. Hamiltonian (c), with parameters $\Omega/\text{V}_{dd}=0.1$ and $\delta/\text{V}_{dd}=1.1$, occurring at the end of stage 3. \justifying Dashed lines mark energies of the ground state (blue) along with the energy of the state with the highest ground state transition rate (yellow), and subsequent states corresponding to the highest transition rates from the preceding state (green,red), while dots of matching color represent off-diagonal matrix elements.}
\end{figure}

Further insight into the behavior of noise in the initial and final stages of the adiabatic state preparation is provided through the analysis of dominant matrix elements that facilitate phase noise excitation. Since the Hamiltonian in Eq.\eqref{Ryd Hamil} is defined in terms of the many-body number operator $\hat{n}=\sum_k ^N\hat{n}_k$, it is useful to work in the rotating frame of the atoms themselves, which see the phase factor in the transverse drive as a stochastic fluctuation in detuning $\delta(t)\rightarrow\delta(t)-\dot{\phi}(t)$, where $\dot{\phi}(t)$ is the instantaneous change in frequency due to noise \cite{Jandura_2022,Fromonteil_2024}.  We achieve this change of frames by applying the unitary transformation $\hat{U}_t=\mathrm{exp}({-i\phi(t)}\sum\hat{n}_k)$, as shown in Appendix \ref{app: Rotating frame}. Figure \ref{fig: Matrix Elements} shows the distribution of transition rates between different many-body energy eigenstates for the Hamiltonian defined in Eq. \eqref{Ryd Hamil} with parameters that occur at three different points throughout stages 2 and 3 of the adiabatic state preparation studied. Each plot shows transitions for the ground state $|E_0\rangle$, along with transitions for the state with the highest transition rate with $|E_0\rangle$ and again with states that have the highest transition rates with the preceding analyzed state. In this way, we trace the most probable excitation path for a population starting in the ground state, getting insight on how excitation dynamics changes throughout the protocol. We begin in Figure \ref{fig: Matrix Elements}(a) with the Hamiltonian occurring in the middle of stage 2, where the dominant transition is that from the ground state $|E_0\rangle$ to the sixth excited eigenstate $|E_6\rangle$ at an energy of $E_6/V_{dd}~\approx1.3$. The remaining cohort of non-negligible transition probabilities is at least an order of magnitude smaller, found at energies similar to $E_6$ and to a lesser degree around $E/V_{dd}~\approx3.3$. As $|E_6\rangle$ becomes excited over time, its transition elements also become relevant, with two dominant excitation channels that excite upward to $|E_{31}\rangle$ with energy $E_{31}/V_{dd}\approx2.7$ and other nearby states, or de-excite back to $|E_0\rangle$ leading to an oscillatory behavior between the two states. This behavior is repeated for $|E_{31}\rangle$. The increase in detuning throughout stage 2 not only decreases the energy gaps between states, making diabatic excitation more likely, but also increases the statistical weight across all off-diagonal elements. Moreover, a high transverse field of $\Omega/V_{dd}=0.6$ induces state mixing, which promotes higher energy transitions as the system in a more non-integrable regime (a discussion on integrability in this system is provided in Appendix \ref{app: Integrability}). Consequently, Stage 2 reconfigures the energetic structure of the many body system to facilitate not only initial phase noise excitation from the ground state, but also subsequent excitations that disperse energy across the entire spectrum. The culmination of these effects can be seen at the transition point between stages 2 and 3 in Figure \ref{fig: Matrix Elements}(b) with laser parameters $\Omega/V_{dd}=0.6$ and $\delta/V_{dd}=1.1$, where we see a dramatic increase in both the density of available states and the magnitude of their respective transition rates. The highlighted excitation path of $|E_0\rangle\rightarrow|E_{13}\rangle\rightarrow|E_{88}\rangle\rightarrow|E_{288}\rangle$ is the most dominant, forming an array of equidistant peaks that can allow fast transfer of energy from the ground state to higher energy sectors. Furthermore, the high number of eigenstates with an elevated transition rate give phase noise further outlet to excite the system at a wider range of frequencies than at any other point throughout the adiabatic process. Transitions for the final Hamiltonian of the protocol are shown in Figure \ref{fig: Matrix Elements}(c), with $\Omega/V_{dd}~=0.1$. Here we see excitation probability from the ground state as well as the higher energy states become strongly suppressed. For $|E_0\rangle$, the most dominant transition has a probability that is two orders of magnitude smaller than the one in Figure \ref{fig: Matrix Elements}(b), suggesting that the ground state is much more robust against noise excitation in the final stages of the ramp. Moreover, the energy eigenvectors that dominated the excitation dynamics in Figure \ref{fig: Matrix Elements}(b) appear as higher excited states, and the equidistant transition peaks of the matrix elements are replaced with transitions for much smaller probabilities that cluster near their respective eigenstate. This suggests that excitation due to laser phase noise is much more constrained in the final stages of the experiment, with available excitation states having smaller transition rates that are all near in energy.

The differences between the last two sets of matrix elements can be explained by the competition between the two laser parameters in the Hamiltonian in Eq. $\eqref{Ryd Hamil}$. As $\Omega/V_{dd}$ decreases, the integrability of the Hamiltonian changes, which has strong implications on the transition element spectrum. This change transitions the system from a non-integrable state with large state mixing to an integrable state that approaches the Ising model \cite{Noh_2021,Deutsch_2018}. Higher integrability leads to an underlying structure in the energy levels, which form clusters separated by large energy gaps (see Figure \ref{fig: SYMplot} in Appendix \ref{app:Reflection}). Such energy gaps, coupled with the absence of matrix elements with distant spectral indices that could facilitate excitation to different energy regions, inhibit the spread of excitation energy throughout the Hilbert space. Of course, the picture is much more complicated in the dynamical setting of the adiabatic state preparation where a time dependent $\Omega(t)$ constantly changes excitation patterns, but Figure \ref{fig: Matrix Elements} along with phase noise evolutions of time-independent Hamiltonians at various stages of the preparation (see Figure \ref{fig: pDE const EVO kappaCMP} in Appendix \ref{AppB: TI phase noise}) provide evidence that phase noise excitation is highly interlinked with the integrability of the system. As the transverse field that breaks integrability decreases, so does the ability of the phase noise to excite the system.

\section{Thermalization \label{Sec 4}}

The presence of phase noise introduces energy into the system. It is then natural to ask if and how this added energy thermalizes across the many-body eigenstates. We define thermalization as a relaxation to states where the values of macroscopic observables become stationary over long periods of time, across widely differing initial conditions, and predictable by the use of statistical mechanics \cite{Deutsch_2018}. The mechanisms underpinning thermalization in quantum systems vary considerably from classical systems in which the process is described through chaotic dynamics in ergodic systems that lead to a Boltzmann distribution across all accessible states over long times. This understanding of thermalization is incompatible with isolated quantum systems, for which dynamical chaos and ergodicity are absent due to the linearity of time evolution and discreteness of the state-space. Instead, quantum thermalization is generally studied by comparing excitation spectra of dynamical quantum systems to a variety of thermal ensembles predicted by statistical mechanics. Such thermalization has been observed in several generic isolated quantum systems \cite{Srednicki_1994,Sengupta_2004,Horoi_1995,Kim_2016,Kim_2018,Khatami_2013}, integrable systems that exhibit a large number of conserved quantities do not reach thermal equilibrium \cite{Rigol_2007,Kinoshita2006AQN,Manmana_2007,Khatami_2013} and instead relax to the generalized Gibbs ensemble \cite{Cassidy_2011,Kollar_2011}. Whether or not thermalization is observed at the end of this adiabatic procedure will therefore depend strongly on the integrability of the Hamiltonian and the choice of observable.

For systems that relax to a thermal equilibrium, the underlying mechanism is most robustly described by the eigenstate thermalization hypothesis (ETH) \cite{Deutsch_2018}. The ETH states that for nonintegrable systems (that do not exhibit many-body localization \cite{Alet_2018,Geraedts_2016}) the expectation values fluctuate negligibly around eigenstates $|E_{i}\rangle$ that are close to the energy of the system. Thus, each individual eigenstate can be thought of as an independent thermal state. In the thermodynamic limit, the resulting state should be constant and predictable by the total energy of the system. The hypothesis also presumes that the observable of interest is `well behaved', in that $\langle E_{i}|\hat{A}|E_{
i}\rangle$ behaves smoothly as a function of eigenstates of the Hamiltonian $|E_i\rangle$, with no sudden discontinuities in the vicinity of the expected energy that would create a sensitivity to slight changes in relaxation dynamics. Given these conditions the ETH predicts that the long time (LT) expectation value of a few body observable $\hat{A}$ with an eigenstate
$|E_{i}\rangle$ of a many-body system Hamiltonian $\hat{H}$
with a corresponding well defined energy $E_{i}$ is equal
to the thermal ensemble $\langle\hat{A}\rangle_{\text{Therm}}(E_{i})$
of $\hat{A}$ at a mean energy $E_{i}$

\begin{equation}
\langle E_{i}|\hat{A}|E_{i}\rangle_{LT}=\langle\hat{A}\rangle_{\text{Therm}}(E_{i}).\label{eq:EigStateHyp}
\end{equation}

\noindent Hence, the ETH provides a framework for comparing expectation values which are directly dependent on initial conditions to thermal ensembles which instead depend only on the total energy of the system. This implies a universal equilibrium for a thermalizing system across many trajectories of equivalent energy. It is then necessary to define exactly what is meant by a long time average $\langle E_{i}|\hat{A}|E_{i}\rangle_{LT}$ in Eq. \eqref{eq:EigStateHyp}. Any dynamics in a quantum system lead to a quantum state $|\psi\rangle=\sum_i c_i |E_i\rangle$ with correlations $c^*_i c_j$ between different excited states $|E_i\rangle$ and $|E_j\rangle$ that cause fluctuations in measured expectation values as a function of time and energy difference 

\begin{equation}
    \langle \psi(t)| \hat{A} |\psi(t)\rangle=\sum_{i,j}e^{-\frac{i}{\hbar}(E_i-E_j)t}c^*_i c_{j}\langle E_i |\hat{A} | E_j \rangle.\label{eq: fluctuations}
\end{equation}

The time it takes for these fluctuations to die down drops with increasing system size, but for small and intermediate systems with large energy gaps this time can be too long to measure or even simulate. However, numerically a long time average can be accessed directly by considering the diagonal ensemble 

 \begin{eqnarray}
\label{Long Time Avg}
\langle \psi| \hat{A} |\psi\rangle_{LT}=\sum_{i}|c_{ii}|^2\langle E_i|\hat{A}|E_i \rangle=\sum_{i}|c_{ii}|^2\hat{A}_{ii},
\end{eqnarray}

\noindent in which we sum over all eigenstates $|E_i\rangle$ and effectively omit correlations by not considering off-diagonal elements in the density matrix of a state. The resulting long time expectation value can then be compared to an appropriate thermal ensemble. 

To understand the ETH fully it is also important to define the thermal ensemble $\langle\hat{A}\rangle_{\text{Therm}}(E_{\alpha})$. In statistical mechanics the thermal equilibrium of a dynamical system is defined in terms of the Boltzmann distribution normalized by a partition function $Z=\sum_i\mathrm{exp}(-\beta E_i)$ where $\beta$ is the thermodynamic temperature of the system. An ensemble of available energy states is then scaled by the thermodynamic temperature $\beta$ which can be calculated using the total energy of the system. Adapting this to the Dirac notation, we define the canonical ensemble as

\begin{equation}
\label{Canonical}
\langle \hat{A} \rangle_{\text{canon}}=\frac{1}{Z}\sum_i \mathrm{exp}(-\beta E_i)\langle E_i|\hat{A}|E_i \rangle,
\end{equation}

\noindent where we sum over all the energy eigenstates $|E_i\rangle$. Previous work has shown that for closed systems such as the one discussed in this paper the canonical ensemble is not always representative of final equilibrium \cite{Rigol_2009}. In such finite systems thermalization occurs when energy in a small neighborhood of eigenstates uses the rest of the eigenstates as a thermal reservoir to dissipate into. Since the reservoir itself is not infinite, we see finite-size effects which become more prevalent at smaller system sizes. Thermalization in isolated quantum systems is typically discussed instead in terms of the microcanonical ensemble which considers only a small energy shell of energies, equally weighted and centered around the expected energy. Nevertheless, this approach is highly dependent on the density of states around the average energy of the system. The exact size of the energy shell should be small in comparison to the energy scale of the system, but large enough to include a statistically significant number of energy eigenstates. Relaxation to the microcanonical ensemble is therefore more readily observed in the middle of the energy spectrum, and for larger system sizes, where gaps between energies narrow down and density of states is higher \cite{Kim_2016}. However, the low energy spectrum which dominates in adiabatic state preparation usually exhibits a much lower density of states, leading to nonsensical microcanonical averages as very few or even no eigenstates fall within the vicinity of a given energy shell. The following section will thus evaluate the ETH using the canonical ensemble.

\subsection{Changing parameter regimes}

\begin{figure}
\hspace*{-5mm}\includegraphics[scale=0.4]{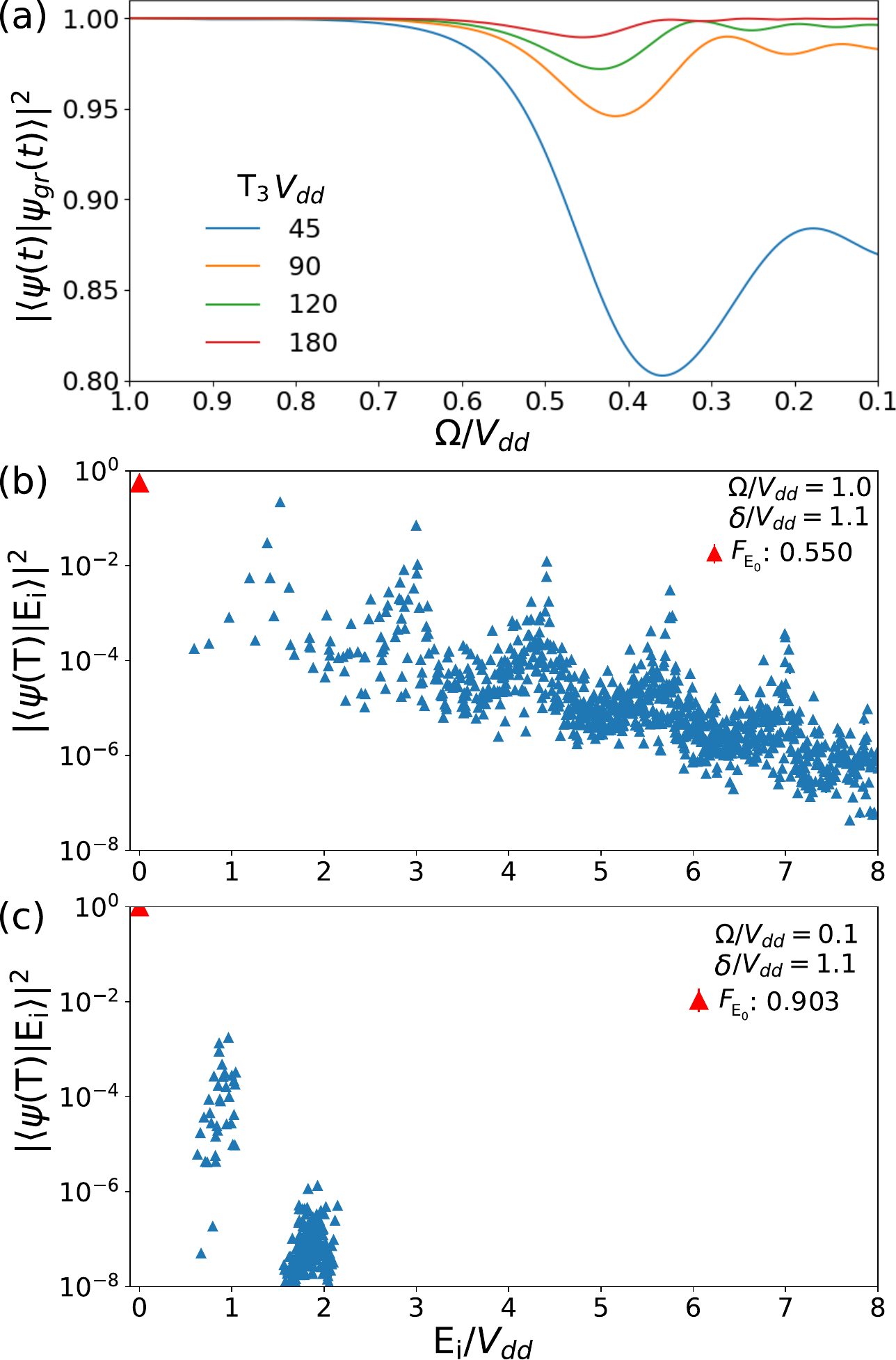}
 \caption{\label{fig:regime change}
 (a) Plot of fidelity against decreasing values of $\Omega/V_{dd}$ for a modified version of the stage 3 adiabatic ramp discussed in Figure \ref{fig:Ryd Adiab Sketch}, now starting from $\Omega/V_{dd}=1.0$. Simulations are performed in an 11 site system with no phase noise, as colored lines show the fidelity for different evolution times with faster ramps incurring more diabatic excitation. (b,c) Overlap of final state and energy eigenstates after a ground state evolution under a time independent Hamiltonian from Eq. \eqref{Ryd Hamil} for a total time of $TV_{dd}=400$ with added phase noise. Plots show simulation for different laser parameters occurring in the modified stage 3 ramp with, detuning set to $\delta/V_{dd}=1.1$ and Rabi frequency set to \textbf{(b)} $\Omega/V_{dd}=1.0$ and \textbf{(c)} $\Omega/V_{dd}=0.1$. \justifying Red triangles mark the ground state fidelity, $\mathrm{F}_{E_0}$, which is also provided in the legend. Error bars plot standard error across 100 time evolutions with independent laser phase noise signals using the same noise profile from Figure \ref{fig:My PSD}, but are too small to be seen on the log scale suggesting consistent excitation.
}%
\end{figure}

To better judge the thermalizing effect of phase noise, it is useful to shift our simulations to a regime in which noise excitation is stronger. In the previous sections, the focus was placed on studying the impact of laser phase noise on high-fidelity adiabatic state preparation. Operating in the $V_{dd}/2\pi = 10~\mathrm{MHz}$ regime kept the characteristic system frequencies well above the dominant noise peak near $0.48~\mathrm{MHz}$ (see Figure \ref{fig:My PSD}), which suppressed excitation and maintained the system close to its instantaneous ground state. In contrast, this section examines the dynamics from the opposite perspective; rather than mitigating phase noise, we exploit its coupling to the system to study how noise can drive energy redistribution and apparent thermalization in an interacting Rydberg chain. For this purpose, we shift to a lower frequency interaction regime of $V_{dd}/2\pi=0.16~\mathrm{MHz}$ that is also studied in Figure \ref{fig: pDE const EVO kappaCMP} in Appendix \ref{AppB: TI phase noise}, where the characteristic transition energies of the Hamiltonian become comparable to the spectral components of the laser noise. In this new regime, preparation of high-fidelity adiabatic states is no longer feasible, as shown in Figure \ref{fig:NOISEPoW and Vdd shift}(b), in which the fidelity of the prepared state decreases to below 20\% even for the smallest system size of 11 sites. Instead, our analysis of thermalization is isolated to stage 3 of the adiabatic ramp, in which phase noise dynamics undergo a dramatic change as the system moves from high to low integrability when the transverse field is reduced. To provide more time for the noise, we prolonge this transition by starting the adiabatic ramp from $\Omega/V_{dd}=1.0$ as opposed to $\Omega/V_{dd}=0.6$ as was the case in Section \ref{Sec 3}.

Figure \ref{fig:regime change} demonstrates the key changes in the dynamics of the modified adiabatic ramp for stage 3. Firstly, Figure \ref{fig:regime change}(a) plots fidelity with instantaneous ground state against the full range of $\Omega/V_{dd}$ for simulations of an 11 site system with no phase noise and for a variety of evolution times. We begin by initializing in the ground state of the initial Hamiltonian of our modified ramp with laser parameters $\Omega/V_{dd}=1.0$ and $\delta/V_{dd}=1.1$. Simulations without laser phase noise reveal that the main loss of fidelity due to diabatic excitation is now shifted to the middle of stage 3, coinciding with the critical region in the ground state energy gap that occurs at $\Omega/V_{dd}\approx0.55$ (see Figure \ref{fig:Ryd Adiab Sketch}(c). Following this, Figures \ref{fig:regime change}(b,c) provide the energy eigenstate overlap of ground state evolved with phase noise under two time independent Hamiltonians that occur at the beginning and end of the adiabatic protocol. We evolve an 11 site system for a long evolution time of $TV_{dd}=400$ to allow enough time for excitation, and observe a stark difference in the amount and nature of phase noise excitation in the two Hamiltonians. Simulations in Figure \ref{fig:regime change}(b) at $\Omega/V_{dd}=1.0$ result in considerable excitation and a final ground state fidelity of $55.0$\%. Moreover, we observe a distinct excitation profile of equidistant peaks corresponding to states with dominant transition rates and high separation in the energy spectrum (see Figure \ref{fig: Matrix Elements}(b) for the same mechanism in $\Omega/V_{dd}=0.6$), which allow a rapid transfer of energy from the ground state to higher energy sectors. However, below $\Omega/V_{dd}=0.6$ the energy landscape begins to undergo dramatic changes as energy states begin to cluster together, and by the time we reach the final Hamiltonian in Figure \ref{fig:regime change}(c) the transition rates are suppressed, especially for eigenstates with large energy gaps (see Figure \ref{fig: SYMplot} in Appendix \ref{app:Reflection}). As a consequence, the final ground state fidelity of time independent Hamiltonian with $\Omega/V_{dd}=0.1$ remains at $90.3\%$, as phase noise finds it much harder to excite the system.

\subsection{Evaluating thermalization}

In the case of adiabatic state preparation with experimentally relevant levels of phase noise, a comparatively small amount of energy is added to the system. This means that the dynamics will be mainly contained in the low energy spectrum where the microcanonical average is not well defined, limiting the analysis to the canonical ensemble. A value for $\beta$ of a system at a given energy can always be calculated by solving the equation 

\begin{equation}
\langle \psi| \hat{H}_R|\psi \rangle=\frac{1}{Z}\sum_i \mathrm{exp}(-\beta E_i)\langle E_i|\hat{H}_R|E_i \rangle,
\end{equation}

\noindent where $|\psi \rangle$ represents the state of the system at the end of the adiabatic state preparation such that the left-hand side of the equation gives the final energy at the end of the procedure averaged over 100 unique realizations. Once a value for $\beta$ is determined it can be used in Eq. \eqref{Canonical} to evaluate arbitrary thermal expectation values at the appropriate energy. Given that phase noise has the potential for exciting across the full energy spectrum it is important to access the full Hilbert space, meaning that we limit ourselves to a system size of 11 sites which can be completely diagonalized.

\begin{figure}
\hspace*{-5mm}\includegraphics[scale=0.4]{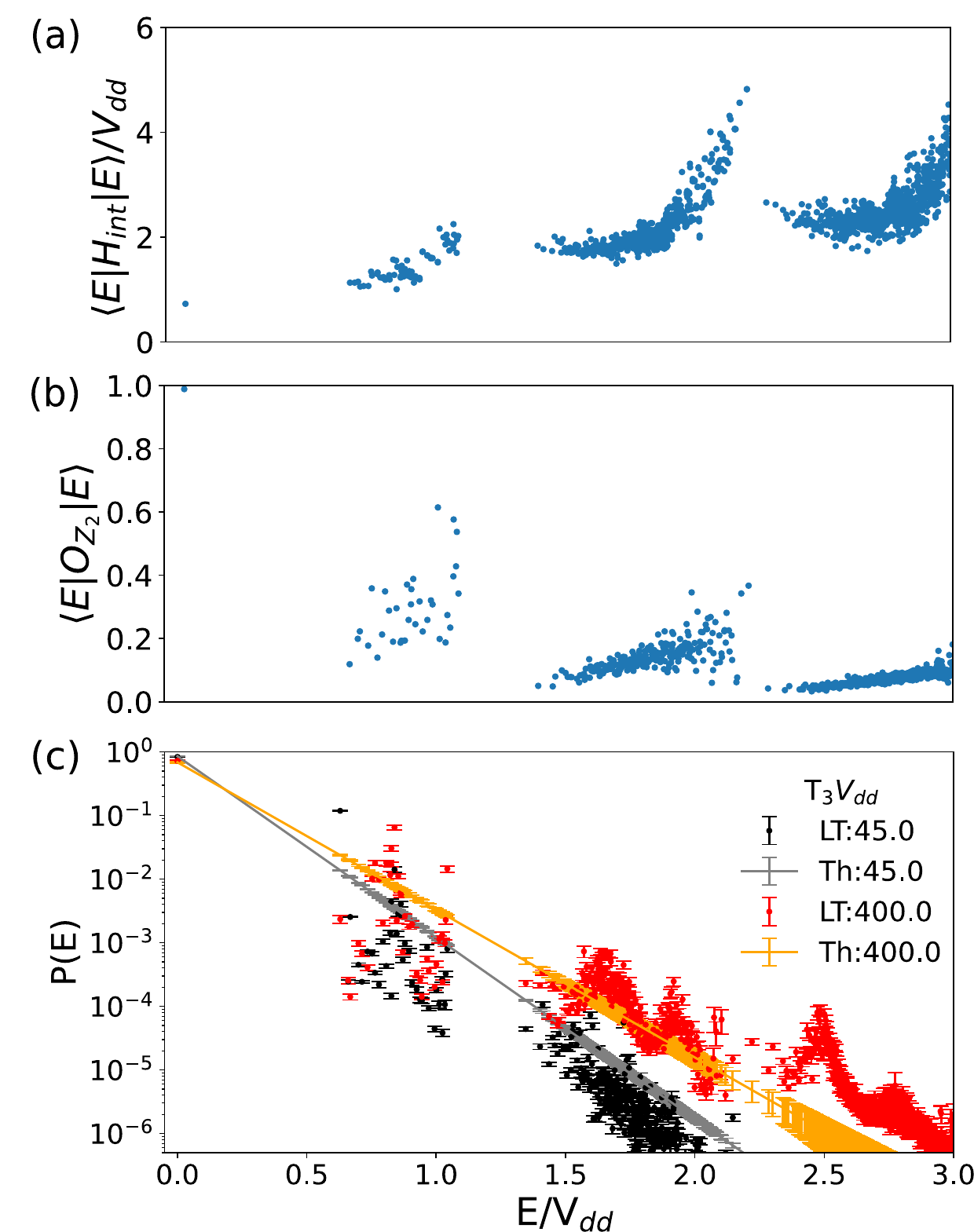}
 \caption{\label{fig:pdist+Hint+SzSz}
 (a) Interaction energy $H_{int}$ defined in Eq. \eqref{H_int} and (b) $\hat{O}_{Z_2}$ ordering parameter as a function of low energy eigenstates of the target Hamiltonian with $\Omega=0.1V_{dd}$ and $\delta=1.1V_{dd}$ in a one dimensional Rydberg spin chain with 11 sites. (c) Long time ensemble (LT) $\langle \Psi|E_i\rangle$, and thermal ensemble (Th) as predicted by the canonical ensemble from Eq. \eqref{Canonical}, for the final step of the adiabatic state preparation with durations $\text{T}_3V_{dd}=45,400$. \justifying Energy is defined with respect to $E_{gr}$, and error bars show standard error across 100 noise realizations.
}%
\end{figure}

Figure \ref{fig:pdist+Hint+SzSz} provides all the distributions required to generate long time and thermal expectation values as a function of energy relative to the ground state. Figures \ref{fig:pdist+Hint+SzSz}(a) and \ref{fig:pdist+Hint+SzSz}(b) show the two observables chosen for the comparison to be the interaction energy $\hat{H}_{int}$ as defined in Eq. \eqref{H_int}, as well as a $\hat{O}_{Z_2}$ order parameter defined in Eq. \eqref{Z2} that measures long range $Z_2$ ordering that is maximally represented in the ground state. Given the relatively small Rabi frequency $\Omega=0.1V_{dd}$ at the end of the state preparation, we are investigating a Hamiltonian that is highly integrable, with structure in energy levels leading to regions of clustered eigenvalues (see Figure \ref{fig: SYMplot} in Appendix \ref{noiseEXC}). Although neither of these observables vary smoothly as a function of energy as prescribed by the ETH, there is considerably less variance in $H_{int}$ compared to the $\hat{O}_{Z_2}$ order parameter meaning that small changes in excitation energy should affect expectation values of the latter more. The distributions as a function of eigenstate energy for $H_{int}$ and the $\hat{O}_{Z_2}$ order parameter are then weighted by the diagonal and canonical ensembles. Figure\ref{fig:pdist+Hint+SzSz}(c) provides example distributions for a short preparation of $T_3V_{dd}=45$ and a much longer preparation at $T_3V_{dd}=400$. Practically speaking, the determining factors of thermalization will be: how well the uniform Boltzmann distribution of the canonical ensemble (with an equivalent many-body energy) acts as a `line of best fit' for the long-time distribution, and, at energies where these distributions disagree, how does the local behavior of the tested observable exacerbate or lessen the impact on the averaged expectation value. Moreover, since we are dealing with ground sate preparation, most of the energy will be held in the low-energy spectrum even after longer evolution times, meaning that the low-energy eigenvalues will have an out-sized impact on the final expectation values. The diagonal ensemble for the shorter $T_3V_{dd}=45$ shows the first excited state having the highest representation, with a majority of higher energy states left largely unexcited and thus with an lower occupation than is predicted by the Boltzmann distribution. For a much longer preparation of $T_3V_{dd}=400$ the first excited state drops far below the occupation of the Boltzmann distribution, as diabatic excitation is subdued. Instead, we see a general rise in occupancy across the entire energy spectrum, with a cohort of low energy eigenstates clustered below $E=1V_{dd}$ from the ground state being most prominent.

\begin{figure}
\includegraphics[scale=0.4]{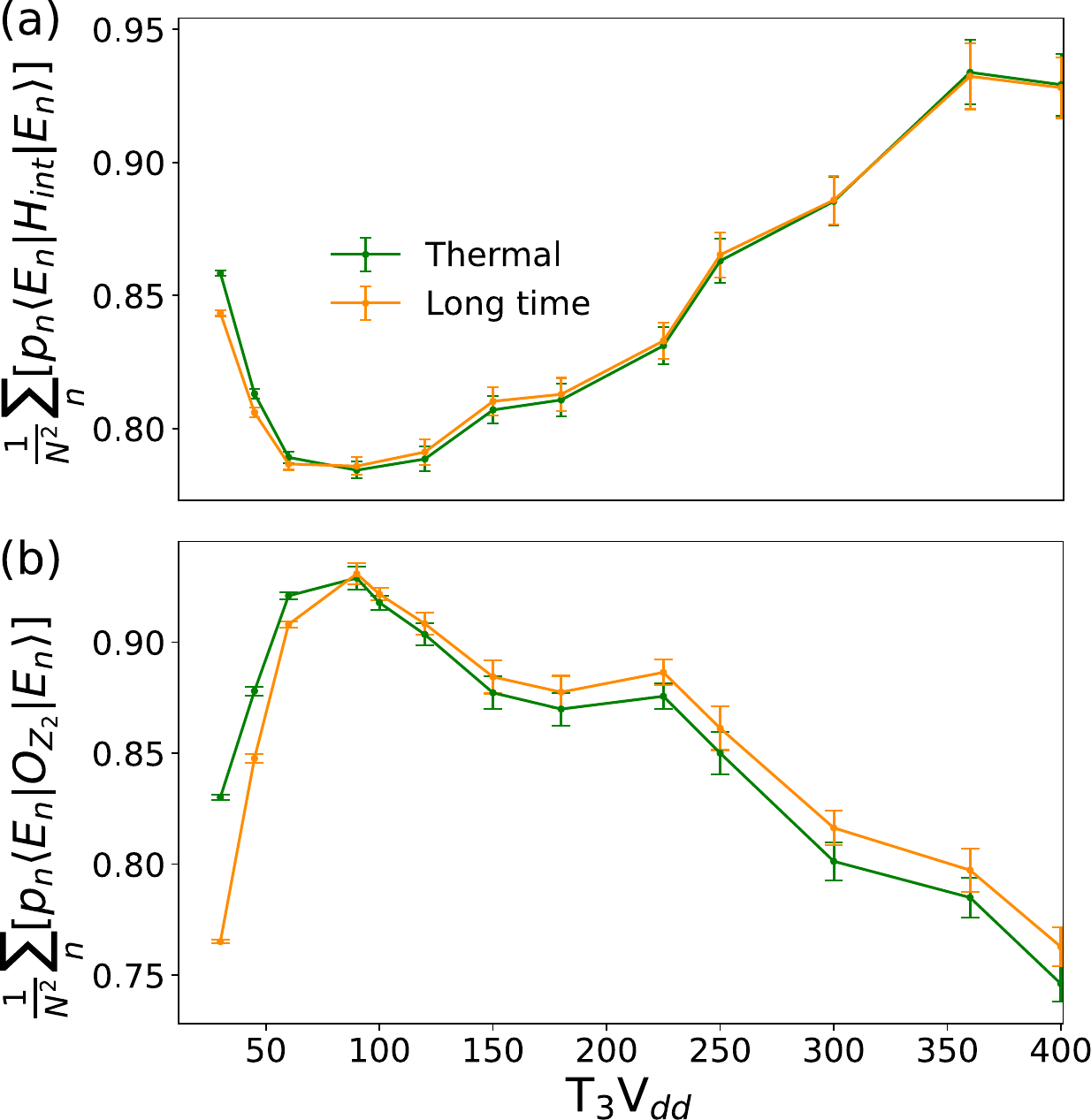}
 \caption{\label{fig:pDE + pTh Expect} Expectation values for (a) interaction energy $H_{int}$ and  (b) $\hat{O}_{Z_2}$ ordering as a function of time taken to complete the third step in the adiabatic state preparation $T_3$. Thermal expectation values (green) that average over the canonical ensemble are compared to long time expectation values (orange) that average over the diagonal ensemble. \justifying Error bars show standard error over 100 preparations with laser phase noise realizations.
 }%
\end{figure}

Averaging over the operator distributions of Figures \ref{fig:pdist+Hint+SzSz}(a) and \ref{fig:pdist+Hint+SzSz}(b) weighted by the long time and thermal ensembles of a particular $T_3$ result in two separate expectation values. Figure \ref{fig:pDE + pTh Expect} provides a comparison between thermal and long time expectation values of $\hat{H}_{int}$ and the order parameter $\hat{O}_{Z_2}$, across a chain of 11 sites, and for a variety of $T_3$. Despite the regime change, the general relationship between both types of expectation value and time $T_3$ is roughly consistent with the results in Figure \ref{fig:noisy Rtime}, as the optimal ramp time when minimal energy is added to the system coincides with the lowest interaction energy in the system and therefore the highest fidelity with the $Z_2$ ordered ground state. For durations shorter than the optimal fidelity, where diabatic excitation dominates, we see a clear divergence between the long time and thermal expectation values. The cause of this divergence can be seen in Figure \ref{fig:pdist+Hint+SzSz}(c), which shows that for the shorter preparation time of $T_3V_{dd}=45$ a majority of energy eigenstates have occupation probabilities below the Boltzmann distribution. However, as the duration of $T_3$ passes its optimum, and laser phase noise begins to dominate excitation, we see a clear convergence of the long time and thermal expectation values. In the case of $H_{int}$ the convergence is strong as soon as $T_3V_{dd}=60$, coinciding with a plateau of minimum energy seen for the same system size of 11 sites in Figure \ref{fig:noisy Rtime}(b). At the optimal $T_3$ there is not enough energy added to the system to differentiate the two expectation values. However, this strong convergence persists as phase noise excitations heat the system. In the case of the $\hat{O}_{Z_2}$ order parameter, a similar divergence of the two expectation values occurs at short $T_3$, along with convergence around the optimal time. For longer $T_3$, the expectation values for this parameter behave more erratically as can be expected given the large variance of the operator as a function of energy. Moreover, the long time expectation value is consistently above the thermal prediction as energy in the long time ensemble is spread over only the ground state and a few energy sectors with high transition rates, meaning that there is always some more occupancy in the perfectly $Z_2$ ordered ground state than would be predicted by the Boltzmann distribution that uniform exponential decay across all eigenstates. Despite this, there is still good convergence between the thermal and long time expectation values even for the $\hat{O}_{Z_2}$ operator.

Analysis of the diagonal ensemble for a ground state evolution of constant Hamiltonians at different $\Omega$ along the simulated adiabatic preparation reveals that phase noise excitation is considerably more prominent in the initial stages of the ramp (see Figure \ref{fig: pDE const EVO kappaCMP} in Appendix \ref{AppB: TI phase noise}). When $\Omega$ is large enough to break integrability, and the phase noise is given enough time, the highly delocalized transitions that dominate excitation form energy peaks across the entire spectrum that remain visible even at the end of the simulation in Figure \ref{fig:pdist+Hint+SzSz}(c). However, by the end of the state preparation when $\Omega$ decreases and the Hamiltonian becomes integrable, these excitation peaks become distorted as energy eigenstates shuffle into tightly packed clusters separated by large energy gaps. There is an overall drop in transition probability as highly delocalized excitations disappear, limiting the energy mobility around the system to the local heating of neighboring eigenstates predicted by Figure \ref{fig: Matrix Elements}(b) and shown in Figure \ref{fig: pDE const EVO kappaCMP}(d). The combination of aggressive delocalized excitation that facilitates rapid energy transfer in the first half of the ramp, and the slow localized heating of this transferred energy in the latter half seems to excite the system in a way such that averaged behavior is lockstep with the uniform predictions of the Boltzmann distribution, albeit the distributions differ considerably. In this way, we can say the adiabatic state preparation facilitates the convergence of the long-time and thermal ensemble. However, when phase noise is left uncontrolled in a non-integrable regime it leads to a diagonal ensemble that rises above the canonical ensemble and does not resemble thermal equilibrium. Similarly, evolving in a Hamiltonian that is fully or close to integrable leads to much less total energy transfer, with strictly localized excitation inhibiting the uniform excitation of the full energy spectrum.

\section{Conclusion \label{Conclusion}}

This study has examined the impact of laser phase noise on adiabatic state preparation in a one-dimensional Rydberg spin chain, focusing on its role in driving excitations and influencing thermalization dynamics. By incorporating phase noise sampled from realistic laser power spectra, the simulations provide a realistic assessment of how coherence errors in the transverse field affect quantum state preparation. The results highlight a competition between diabatic and noise-induced excitations, leading to an optimal ramp duration that minimizes total energy absorption and improves final fidelities. The role of noise resonance and noise scaling is also examined, showing that phase noise is most disruptive when its dominant frequency components are resonant with key system excitations. The results also demonstrate that increasing the overall noise strength leads to a rapid decline in performance, underscoring the need for noise suppression techniques in scalable neutral atom platforms. Additionally,  shifting the system's operational frequency away from the most affected regions by changing the interaction strength between sites improves the fidelity of the state preparation.

Excitation due to phase noise is found to depend strongly on the integrability of the system, with non-integrable regimes seen in the simulated adiabatic ramp facilitating energy delocalization, while integrable regimes lead to energy clustering and large spectral gaps which limit the energy transfer to low probability localized excitations. The interplay of these two regimes has direct implications on how energy is dissipated across the system, as we investigate whether this resembles many-body thermalization. Analysis of long-time expectation values indicates that noise-driven excitations can, under certain conditions, lead to thermalization consistent with the eigenstate thermalization hypothesis (ETH). The observed convergence between diagonal and thermal ensembles suggests that in these systems the excitations generate correlation functions that are close to what would be expected in a thermal ensemble. However, we should note that in important parts of the parameter space, these systems are approximately integrable and will generally lead to deviations from thermalization – even more so when the noise strength is high enough to induce excessive heating. In contrast, in strictly integrable regimes, the absence of delocalized excitation channels restricts energy transfer across the spectrum, and for higher energy states in particular, also preventing the system from reproducing thermal predictions.

Our results have broad implications for quantum simulation and computation with Rydberg atom arrays, especially as system sizes increase and noise mitigation becomes a growing challenge. Future research in this topic could explore a broader range of phase noise spectra \cite{Jiang_2023}, higher-dimensional Rydberg geometries, as well as observing the effects of phase noise on different Rydberg accessible interacting Hamiltonians. 
%A deeper study of how the phase noise excitation mechanism scales with system size is warranted; however, this type of analysis remains computationally challenging as access to the entire Hilbert space is needed for proper evaluation. 
Also of interest is a parallel investigation of phase noise in the Van der Waals Rydberg interaction regime in which interactions scale as $1/r^6$ as opposed to the $1/r^3$ scaling of dipole-dipole interactions, and has become common in experimental setups in recent years \cite{Labuhn_2016,Zhang_2020,Zeybek_2023,de_Oliveira_2025}. %By addressing these challenges, the study aims to contribute to the development of more robust and precise neutral atom platforms for quantum technologies.

\section*{Aknowledgements}
We are grateful to Sebastian Schmidt, Sridevi Kuriyattil, and the members of the SQUARE project team for helpful discussions. This work is supported by the EPSRC (Grant No. EP/T005386/1) and M Squared Lasers Ltd. The data presented in this work are available at \cite{pureDOI_TK}.

%BtIBSTYLE
\bibliography{Q_references}% Produces the bibliography via BibTeX.

@article{itensor,
	title={{The ITensor Software Library for Tensor Network Calculations}},
	author={Matthew Fishman and Steven R. White and E. Miles Stoudenmire},
	journal={SciPost Phys. Codebases},
	pages={4},
	year={2022},
	publisher={SciPost},
	doi={10.21468/SciPostPhysCodeb.4},
	url={https://scipost.org/10.21468/SciPostPhysCodeb.4}
}

@article{Ebadi2022,
author = {S. Ebadi  and A. Keesling  and M. Cain  and T. T. Wang  and H. Levine  and D. Bluvstein  and G. Semeghini  and A. Omran  and J.-G. Liu  and R. Samajdar  and X.-Z. Luo  and B. Nash  and X. Gao  and B. Barak  and E. Farhi  and S. Sachdev  and N. Gemelke  and L. Zhou  and S. Choi  and H. Pichler  and S.-T. Wang  and M. Greiner  and V. Vuletić  and M. D. Lukin },
title = {Quantum optimization of maximum independent set using Rydberg atom arrays},
journal = {Science},
volume = {376},
number = {6598},
pages = {1209-1215},
year = {2022},
doi = {10.1126/science.abo6587},
URL = {https://www.science.org/doi/abs/10.1126/science.abo6587},
eprint = {https://www.science.org/doi/pdf/10.1126/science.abo6587}}

@article{Ravets2014,
  author    = {Ravets, Sylvain and Labuhn, Henning and Barredo, Daniel and B{\'e}guin, Lucas and Lahaye, Thierry and Browaeys, Antoine},
  title     = {Coherent dipole--dipole coupling between two single {R}ydberg atoms at an electrically-tuned {F}{\"o}rster resonance},
  journal   = {Nature Physics},
  year      = {2014},
  volume    = {10},
  number    = {12},
  pages     = {914--917},
  month     = {dec},
  doi       = {10.1038/nphys3119},
  url       = {https://doi.org/10.1038/nphys3119},
  publisher = {Nature Publishing Group}
}

@article{Saffman2010,
  title = {Quantum information with Rydberg atoms},
  author = {Saffman, M. and Walker, T. G. and M\o{}lmer, K.},
  journal = {Rev. Mod. Phys.},
  volume = {82},
  issue = {3},
  pages = {2313--2363},
  numpages = {0},
  year = {2010},
  month = {Aug},
  publisher = {American Physical Society},
  doi = {10.1103/RevModPhys.82.2313},
  url = {https://link.aps.org/doi/10.1103/RevModPhys.82.2313}
}

@article{Schach2010, 
doi = {10.1088/1367-2630/12/10/103044}, 
url = {https://dx.doi.org/10.1088/1367-2630/12/10/103044}, 
year = {2010}, 
month = {oct}, 
publisher = {}, 
volume = {12}, 
number = {10}, 
pages = {103044}, 
author = {Schachenmayer, J and Lesanovsky, I and Micheli, A and Daley, A J}, 
title = {Dynamical crystal creation with polar molecules or Rydberg atoms in optical lattices}, 
journal = {New Journal of Physics} 
}

@article{Sengupta_2004,
   title={Quench dynamics across quantum critical points},
   volume={69},
   ISSN={1094-1622},
   url={http://dx.doi.org/10.1103/PhysRevA.69.053616},
   DOI={10.1103/physreva.69.053616},
   number={5},
   journal={Physical Review A},
   publisher={American Physical Society (APS)},
   author={Sengupta, K. and Powell, Stephen and Sachdev, Subir},
   year={2004},
   month=may }

@article{Rigol_2007,
   title={Relaxation in a Completely Integrable Many-Body Quantum System: AnAb InitioStudy of the Dynamics of the Highly Excited States of 1D Lattice Hard-Core Bosons},
   volume={98},
   ISSN={1079-7114},
   url={http://dx.doi.org/10.1103/PhysRevLett.98.050405},
   DOI={10.1103/physrevlett.98.050405},
   number={5},
   journal={Physical Review Letters},
   publisher={American Physical Society (APS)},
   author={Rigol, Marcos and Dunjko, Vanja and Yurovsky, Vladimir and Olshanii, Maxim},
   year={2007},
   month=feb }

@article{Rigol_2009,
   title={Breakdown of Thermalization in Finite One-Dimensional Systems},
   volume={103},
   ISSN={1079-7114},
   url={http://dx.doi.org/10.1103/PhysRevLett.103.100403},
   DOI={10.1103/physrevlett.103.100403},
   number={10},
   journal={Physical Review Letters},
   publisher={American Physical Society (APS)},
   author={Rigol, Marcos},
   year={2009},
   month=sep }

@article{Kinoshita2006AQN,
  title={A quantum Newton's cradle},
  author={Toshiya Kinoshita and Trevor Wenger and David S. Weiss},
  journal={Nature},
  year={2006},
  volume={440},
  pages={900-903},
  url={https://api.semanticscholar.org/CorpusID:20114248}
}

@article{Geraedts_2016,
  title = {Many-body localization and thermalization: Insights from the entanglement spectrum},
  author = {Geraedts, Scott D. and Nandkishore, Rahul and Regnault, Nicolas},
  journal = {Phys. Rev. B},
  volume = {93},
  issue = {17},
  pages = {174202},
  numpages = {18},
  year = {2016},
  month = {May},
  publisher = {American Physical Society},
  doi = {10.1103/PhysRevB.93.174202},
  url = {https://link.aps.org/doi/10.1103/PhysRevB.93.174202}
}

@article{Srednicki_1994,
   title={Chaos and quantum thermalization},
   volume={50},
   ISSN={1095-3787},
   url={http://dx.doi.org/10.1103/PhysRevE.50.888},
   DOI={10.1103/physreve.50.888},
   number={2},
   journal={Physical Review E},
   publisher={American Physical Society (APS)},
   author={Srednicki, Mark},
   year={1994},
   month=aug, pages={888–901} }

@article{Cassidy_2011,
   title={Generalized Thermalization in an Integrable Lattice System},
   volume={106},
   ISSN={1079-7114},
   url={http://dx.doi.org/10.1103/PhysRevLett.106.140405},
   DOI={10.1103/physrevlett.106.140405},
   number={14},
   journal={Physical Review Letters},
   publisher={American Physical Society (APS)},
   author={Cassidy, Amy C. and Clark, Charles W. and Rigol, Marcos},
   year={2011},
   month=apr }

@article{Manmana_2007,
  title = {Strongly Correlated Fermions after a Quantum Quench},
  author = {Manmana, S. R. and Wessel, S. and Noack, R. M. and Muramatsu, A.},
  journal = {Phys. Rev. Lett.},
  volume = {98},
  issue = {21},
  pages = {210405},
  numpages = {4},
  year = {2007},
  month = {May},
  publisher = {American Physical Society},
  doi = {10.1103/PhysRevLett.98.210405},
  url = {https://link.aps.org/doi/10.1103/PhysRevLett.98.210405}
}

@article{Kollar_2011,
   title={Generalized Gibbs ensemble prediction of prethermalization plateaus and their relation to nonthermal steady states in integrable systems},
   volume={84},
   ISSN={1550-235X},
   url={http://dx.doi.org/10.1103/PhysRevB.84.054304},
   DOI={10.1103/physrevb.84.054304},
   number={5},
   journal={Physical Review B},
   publisher={American Physical Society (APS)},
   author={Kollar, Marcus and Wolf, F. Alexander and Eckstein, Martin},
   year={2011},
   month=aug }

@article{Alet_2018,
   title={Many-body localization: An introduction and selected topics},
   volume={19},
   ISSN={1878-1535},
   url={http://dx.doi.org/10.1016/j.crhy.2018.03.003},
   DOI={10.1016/j.crhy.2018.03.003},
   number={6},
   journal={Comptes Rendus. Physique},
   publisher={Cellule MathDoc/Centre Mersenne},
   author={Alet, Fabien and Laflorencie, Nicolas},
   year={2018},
   month=apr, pages={498–525} }

@article{Horoi_1995,
  title = {Chaos vs Thermalization in the Nuclear Shell Model},
  author = {Horoi, Mihai and Zelevinsky, Vladimir and Brown, B. Alex},
  journal = {Phys. Rev. Lett.},
  volume = {74},
  issue = {26},
  pages = {5194--5197},
  numpages = {0},
  year = {1995},
  month = {Jun},
  publisher = {American Physical Society},
  doi = {10.1103/PhysRevLett.74.5194},
  url = {https://link.aps.org/doi/10.1103/PhysRevLett.74.5194}
}

@article{Deutsch_2018,
doi = {10.1088/1361-6633/aac9f1},
url = {https://dx.doi.org/10.1088/1361-6633/aac9f1},
year = {2018},
month = {jul},
publisher = {IOP Publishing},
volume = {81},
number = {8},
pages = {082001},
author = {Joshua M Deutsch},
title = {Eigenstate thermalization hypothesis},
journal = {Reports on Progress in Physics}
}

@article{Johnson2008,
  title = {Rabi Oscillations between Ground and Rydberg States with Dipole-Dipole Atomic Interactions},
  author = {Johnson, T. A. and Urban, E. and Henage, T. and Isenhower, L. and Yavuz, D. D. and Walker, T. G. and Saffman, M.},
  journal = {Phys. Rev. Lett.},
  volume = {100},
  issue = {11},
  pages = {113003},
  numpages = {4},
  year = {2008},
  month = {Mar},
  publisher = {American Physical Society},
  doi = {10.1103/PhysRevLett.100.113003},
  url = {https://link.aps.org/doi/10.1103/PhysRevLett.100.113003}
}

@article{Lukin2001,
  title = {Dipole Blockade and Quantum Information Processing in Mesoscopic Atomic Ensembles},
  author = {Lukin, M. D. and Fleischhauer, M. and Cote, R. and Duan, L. M. and Jaksch, D. and Cirac, J. I. and Zoller, P.},
  journal = {Phys. Rev. Lett.},
  volume = {87},
  issue = {3},
  pages = {037901},
  numpages = {4},
  year = {2001},
  month = {Jun},
  publisher = {American Physical Society},
  doi = {10.1103/PhysRevLett.87.037901},
  url = {https://link.aps.org/doi/10.1103/PhysRevLett.87.037901}
}

@article{Barredo2016,
author = {Daniel Barredo  and Sylvain de Léséleuc  and Vincent Lienhard  and Thierry Lahaye  and Antoine Browaeys },
title = {An atom-by-atom assembler of defect-free arbitrary two-dimensional atomic arrays},
journal = {Science},
volume = {354},
number = {6315},
pages = {1021-1023},
year = {2016},
doi = {10.1126/science.aah3778},
URL = {https://www.science.org/doi/abs/10.1126/science.aah3778},
eprint = {https://www.science.org/doi/pdf/10.1126/science.aah3778}}

@article{Morgado2020,
author = {Morgado, Manuel and Whitlock, Shannon},
year = {2021},
month = {06},
pages = {023501},
title = {Quantum simulation and computing with Rydberg-interacting qubits},
volume = {3},
journal = {AVS Quantum Science},
url={https://api.semanticscholar.org/CorpusID:226254232},
doi = {10.1116/5.0036562}
}

@article{Kim_2016,
  title = {Testing whether all eigenstates obey the eigenstate thermalization hypothesis},
  author = {Kim, Hyungwon and Ikeda, Tatsuhiko N. and Huse, David A.},
  journal = {Phys. Rev. E},
  volume = {90},
  issue = {5},
  pages = {052105},
  numpages = {8},
  year = {2014},
  month = {Nov},
  publisher = {American Physical Society},
  doi = {10.1103/PhysRevE.90.052105},
  url = {https://link.aps.org/doi/10.1103/PhysRevE.90.052105}
}

@article{Kim_2018,
   title={Detailed Balance of Thermalization Dynamics in Rydberg-Atom Quantum Simulators},
   volume={120},
   ISSN={1079-7114},
   url={http://dx.doi.org/10.1103/PhysRevLett.120.180502},
   DOI={10.1103/physrevlett.120.180502},
   number={18},
   journal={Physical Review Letters},
   publisher={American Physical Society (APS)},
   author={Kim, Hyosub and Park, YeJe and Kim, Kyungtae and Sim, H.-S. and Ahn, Jaewook},
   year={2018},
   month=may }

@article{Khatami_2013,
   title={Fluctuation-Dissipation Theorem in an Isolated System of Quantum Dipolar Bosons after a Quench},
   volume={111},
   ISSN={1079-7114},
   url={http://dx.doi.org/10.1103/PhysRevLett.111.050403},
   DOI={10.1103/physrevlett.111.050403},
   number={5},
   journal={Physical Review Letters},
   publisher={American Physical Society (APS)},
   author={Khatami, Ehsan and Pupillo, Guido and Srednicki, Mark and Rigol, Marcos},
   year={2013},
   month=jul }

@article{Henriet2020,
  doi = {10.22331/q-2020-09-21-327},
  url = {https://doi.org/10.22331/q-2020-09-21-327},
  title = {Quantum computing with neutral atoms},
  author = {Henriet, Lo{\"{i}}c and Beguin, Lucas and Signoles, Adrien and Lahaye, Thierry and Browaeys, Antoine and Reymond, Georges-Olivier and Jurczak, Christophe},
  journal = {{Quantum}},
  issn = {2521-327X},
  publisher = {{Verein zur F{\"{o}}rderung des Open Access Publizierens in den Quantenwissenschaften}},
  volume = {4},
  pages = {327},
  month = sep,
  year = {2020}
}

@article{Weber2017,
author = {Weber, Sebastian and Tresp, Christoph and Menke, Henri and Urvoy, Alban and Firstenberg, Ofer and B{\"{u}}chler, Hans Peter and Hofferberth, Sebastian},
doi = {10.1088/1361-6455/aa743a},
issn = {13616455},
journal = {J. Phys. B},
number = {13},
publisher = {IOP Publishing},
title = {{Calculation of Rydberg interaction potentials}},
volume = {50},
year = {2017}
}

@article{Ebadi2020,
  author    = {Sepehr Ebadi and Tout T. Wang and Harry Levine and Alexander Keesling and Giulia Semeghini and Ahmed Omran and Dolev Bluvstein and Rhine Samajdar and Hannes Pichler and Wen Wei Ho and Soonwon Choi and Subir Sachdev and Markus Greiner and Vladan Vuletić and Mikhail D. Lukin},
  title     = {Quantum phases of matter on a 256-atom programmable quantum simulator},
  journal   = {Nature},
  year      = {2021},
  volume    = {595},
  number    = {7866},
  pages     = {227--232},
  doi       = {10.1038/s41586-021-03582-4},
  url       = {https://doi.org/10.1038/s41586-021-03582-4},
  issn      = {1476-4687}
}

@article{Levine2018,
  title = {High-Fidelity Control and Entanglement of Rydberg-Atom Qubits},
  author = {Levine, Harry and Keesling, Alexander and Omran, Ahmed and Bernien, Hannes and Schwartz, Sylvain and Zibrov, Alexander S. and Endres, Manuel and Greiner, Markus and Vuleti\ifmmode \acute{c}\else \'{c}\fi{}, Vladan and Lukin, Mikhail D.},
  journal = {Phys. Rev. Lett.},
  volume = {121},
  issue = {12},
  pages = {123603},
  numpages = {6},
  year = {2018},
  month = {Sep},
  publisher = {American Physical Society},
  doi = {10.1103/PhysRevLett.121.123603},
  url = {https://link.aps.org/doi/10.1103/PhysRevLett.121.123603}
}

@article{Leseleuc17,
  title = {Optical Control of the Resonant Dipole-Dipole Interaction between Rydberg Atoms},
  author = {de L\'es\'eleuc, Sylvain and Barredo, Daniel and Lienhard, Vincent and Browaeys, Antoine and Lahaye, Thierry},
  journal = {Phys. Rev. Lett.},
  volume = {119},
  issue = {5},
  pages = {053202},
  numpages = {6},
  year = {2017},
  publisher = {American Physical Society},
  doi = {10.1103/PhysRevLett.119.053202},
  url = {https://link.aps.org/doi/10.1103/PhysRevLett.119.053202}
}

@article{Browaeys_2016,
doi = {10.1088/0953-4075/49/15/152001},
url = {https://dx.doi.org/10.1088/0953-4075/49/15/152001},
year = {2016},
publisher = {IOP Publishing},
volume = {49},
number = {15},
pages = {152001},
author = {Browaeys, Antoine and Barredo, Daniel and Lahaye, Thierry},
title = {Experimental investigations of dipole–dipole interactions between a few Rydberg atoms},
journal = {Journal of Physics B: Atomic, Molecular and Optical Physics}
}

@article{Ates_2008, 
doi = {10.1088/1367-2630/10/4/045030}, 
url = {https://dx.doi.org/10.1088/1367-2630/10/4/045030}, 
year = {2008}, 
journal = {New Journal of Physics},
publisher = {}, 
volume = {10}, 
number = {4}, 
pages = {045030}, 
author = {Ates, C and Eisfeld, A and Rost, J M}, 
title = {Motion of Rydberg atoms induced by resonant dipoledipole interactions that trigger the energy transfer between two energetically close Rydberg states. How and if the atoms move depends on their initial arrangement as well as on the initial electronic excitation. Using a mixed quantum/classical propagation scheme, we obtain the trajectories and kinetic energies of atoms, initially arranged in a regular chain and prepared in excitonic eigenstates. The influence of the off-diagonal disorder on the motion of the atoms is examined and it is shown that irregularity in the arrangement of the atoms can lead to an acceleration of the nuclear dynamics.} 
}

@article{Leseleuc18,
  title = {Analysis of imperfections in the coherent optical excitation of single atoms to Rydberg states},
  author = {de L\'es\'eleuc, Sylvain and Barredo, Daniel and Lienhard, Vincent and Browaeys, Antoine and Lahaye, Thierry},
  journal = {Phys. Rev. A},
  volume = {97},
  issue = {5},
  pages = {053803},
  numpages = {9},
  year = {2018},
  month = {May},
  publisher = {American Physical Society},
  doi = {10.1103/PhysRevA.97.053803},
  url = {https://link.aps.org/doi/10.1103/PhysRevA.97.053803}
}

@article{Petrosyan_2016,
   title={On the adiabatic preparation of spatially-ordered Rydberg excitations of atoms in a one-dimensional optical lattice by laser frequency sweeps},
   volume={49},
   ISSN={1361-6455},
   url={http://dx.doi.org/10.1088/0953-4075/49/8/084003},
   DOI={10.1088/0953-4075/49/8/084003},
   number={8},
   journal={Journal of Physics B: Atomic, Molecular and Optical Physics},
   publisher={IOP Publishing},
   author={Petrosyan, David and Mølmer, Klaus and Fleischhauer, Michael},
   year={2016},
   month=apr, pages={084003} }

@article{Petrosyan2017,
  title = {High-fidelity Rydberg quantum gate via a two-atom dark state},
  author = {Petrosyan, David and Motzoi, Felix and Saffman, Mark and M\o{}lmer, Klaus},
  journal = {Phys. Rev. A},
  volume = {96},
  issue = {4},
  pages = {042306},
  numpages = {9},
  year = {2017},
  month = {Oct},
  publisher = {American Physical Society},
  doi = {10.1103/PhysRevA.96.042306},
  url = {https://link.aps.org/doi/10.1103/PhysRevA.96.042306}
}

@article{D_Alessio_2015,
   title={Dynamical preparation of Floquet Chern insulators},
   volume={6},
   ISSN={2041-1723},
   url={http://dx.doi.org/10.1038/ncomms9336},
   DOI={10.1038/ncomms9336},
   number={1},
   journal={Nature Communications},
   publisher={Springer Science and Business Media LLC},
   author={D’Alessio, Luca and Rigol, Marcos},
   year={2015},
   month=oct }

@ARTICLE{TK95,
       author = {{Timmer}, J. and {K{\"o}nig}, M.},
        title = "{On generating power law noise.}",
      journal = {Astronomy \& Astrophysics},
         year = {1995},
        month = {03},
       volume = {300},
        pages = {707},
       url = {https://ui.adsabs.harvard.edu/abs/1995A&A...300..707T}
}

@article{Bernien_2017,
   title={Probing many-body dynamics on a 51-atom quantum simulator},
   volume={551},
   ISSN={1476-4687},
   url={http://dx.doi.org/10.1038/nature24622},
   DOI={10.1038/nature24622},
   number={7682},
   journal={Nature},
   publisher={Springer Science and Business Media LLC},
   author={Bernien, Hannes and Schwartz, Sylvain and Keesling, Alexander and Levine, Harry and Omran, Ahmed and Pichler, Hannes and Choi, Soonwon and Zibrov, Alexander S. and Endres, Manuel and Greiner, Markus and Vuletić, Vladan and Lukin, Mikhail D.},
   year={2017},
   month=nov, pages={579–584} }

@article{Jiang_2023,
  title = {Sensitivity of quantum gate fidelity to laser phase and intensity noise},
  author = {Jiang, X. and Scott, J. and Friesen, Mark and Saffman, M.},
  journal = {Phys. Rev. A},
  volume = {107},
  issue = {4},
  pages = {042611},
  numpages = {23},
  year = {2023},
  month = {Apr},
  publisher = {American Physical Society},
  doi = {10.1103/PhysRevA.107.042611},
  url = {https://link.aps.org/doi/10.1103/PhysRevA.107.042611}
}

@article{Pelegrí_2022,
doi = {10.1088/2058-9565/ac823a},
url = {https://dx.doi.org/10.1088/2058-9565/ac823a},
year = {2022},
month = {aug},
publisher = {IOP Publishing},
volume = {7},
number = {4},
pages = {045020},
author = {G Pelegrí and A J Daley and J D Pritchard},
title = {High-fidelity multiqubit Rydberg gates via two-photon adiabatic rapid passage},
journal = {Quantum Science and Technology}
}

@article{Labuhn_2016,
   title={Tunable two-dimensional arrays of single Rydberg atoms for realizing quantum Ising models},
   volume={534},
   ISSN={1476-4687},
   url={http://dx.doi.org/10.1038/nature18274},
   DOI={10.1038/nature18274},
   number={7609},
   journal={Nature},
   publisher={Springer Science and Business Media LLC},
   author={Labuhn, Henning and Barredo, Daniel and Ravets, Sylvain and de Léséleuc, Sylvain and Macrì, Tommaso and Lahaye, Thierry and Browaeys, Antoine},
   year={2016},
   month=jun, pages={667–670} }

@article{Pohl_2010_CrystallineExperiment,
  title = {Dynamical Crystallization in the Dipole Blockade of Ultracold Atoms},
  author = {Pohl, T. and Demler, E. and Lukin, M. D.},
  journal = {Phys. Rev. Lett.},
  volume = {104},
  issue = {4},
  pages = {043002},
  numpages = {4},
  year = {2010},
  month = {Jan},
  publisher = {American Physical Society},
  doi = {10.1103/PhysRevLett.104.043002},
  url = {https://link.aps.org/doi/10.1103/PhysRevLett.104.043002}
}

@article{Glaetzle_2017,
   title={A coherent quantum annealer with Rydberg atoms},
   volume={8},
   ISSN={2041-1723},
   url={http://dx.doi.org/10.1038/ncomms15813},
   DOI={10.1038/ncomms15813},
   number={1},
   journal={Nature Communications},
   publisher={Springer Science and Business Media LLC},
   author={Glaetzle, A. W. and van Bijnen, R. M. W. and Zoller, P. and Lechner, W.},
   year={2017},
   month=jun }

@article{Kadowaki_1998,
   title={Quantum annealing in the transverse Ising model},
   volume={58},
   ISSN={1095-3787},
   url={http://dx.doi.org/10.1103/PhysRevE.58.5355},
   DOI={10.1103/physreve.58.5355},
   number={5},
   journal={Physical Review E},
   publisher={American Physical Society (APS)},
   author={Kadowaki, Tadashi and Nishimori, Hidetoshi},
   year={1998},
   month=nov, pages={5355–5363} }

@article{Schaub_2015,
  author    = {Peter Schauß and Johannes Zeiher and Takeshi Fukuhara and Stefan Hild and Marc Cheneau and Tommaso Macrì and Thomas Pohl and Immanuel Bloch and Christian Gross},
  title     = {Crystallization in Ising quantum magnets},
  journal   = {Science},
  year      = {2015},
  volume    = {347},
  number    = {6229},
  pages     = {1455--1458},
  doi       = {10.1126/science.1258351},
  issn      = {0036-8075},
  eissn     = {1095-9203},
  url       = {https://doi.org/10.1126/science.1258351}
}

@article{Bluvstein_2022,
   title={A quantum processor based on coherent transport of entangled atom arrays},
   volume={604},
   ISSN={1476-4687},
   url={http://dx.doi.org/10.1038/s41586-022-04592-6},
   DOI={10.1038/s41586-022-04592-6},
   number={7906},
   journal={Nature},
   publisher={Springer Science and Business Media LLC},
   author={Bluvstein, Dolev and Levine, Harry and Semeghini, Giulia and Wang, Tout T. and Ebadi, Sepehr and Kalinowski, Marcin and Keesling, Alexander and Maskara, Nishad and Pichler, Hannes and Greiner, Markus and Vuletić, Vladan and Lukin, Mikhail D.},
   year={2022},
   month=apr, pages={451–456} }

@article{Schmid_2019,
   title={Simple phase noise measurement scheme for cavity-stabilized laser systems},
   volume={44},
   ISSN={1539-4794},
   url={http://dx.doi.org/10.1364/OL.44.002709},
   DOI={10.1364/ol.44.002709},
   number={11},
   journal={Optics Letters},
   publisher={Optica Publishing Group},
   author={Schmid, Fabian and Weitenberg, Johannes and Hänsch, Theodor W. and Udem, Thomas and Ozawa, Akira},
   year={2019},
   month=may, pages={2709} }

@article{Kim2024,
  author    = {Kangheun Kim and Minhyuk Kim and Juyoung Park and Andrew Byun and Jaewook Ahn},
  title     = {Quantum computing dataset of maximum independent set problem on king lattice of over hundred Rydberg atoms},
  journal   = {Scientific Data},
  year      = {2024},
  volume    = {11},
  number    = {1},
  pages     = {111},
  doi       = {10.1038/s41597-024-02926-9},
  url       = {https://doi.org/10.1038/s41597-024-02926-9},
  issn      = {2052-4463}
}

@misc{day2021,
  author    = {Matthew L. Day and Pei Jiang Low and Brendan White and Rajibul Islam and Crystal Senko},
  title     = {Limits on atomic qubit control from laser noise},
  journal   = {npj Quantum Information},
  year      = {2022},
  volume    = {8},
  number    = {1},
  pages     = {72},
  doi       = {10.1038/s41534-022-00586-4},
  url       = {https://doi.org/10.1038/s41534-022-00586-4},
  issn      = {2056-6387}
}

@article{Fromonteil_2024,
   title={Hamilton-Jacobi-Bellman equations for Rydberg-blockade processes},
   volume={6},
   ISSN={2643-1564},
   url={http://dx.doi.org/10.1103/PhysRevResearch.6.033333},
   DOI={10.1103/physrevresearch.6.033333},
   number={3},
   journal={Physical Review Research},
   publisher={American Physical Society (APS)},
   author={Fromonteil, Charles and Tricarico, Roberto and Cesa, Francesco and Pichler, Hannes},
   year={2024},
   month=sep }

@article{Jandura_2022,
   title={Time-Optimal Two- and Three-Qubit Gates for Rydberg Atoms},
   volume={6},
   ISSN={2521-327X},
   url={http://dx.doi.org/10.22331/q-2022-05-13-712},
   DOI={10.22331/q-2022-05-13-712},
   journal={Quantum},
   publisher={Verein zur Forderung des Open Access Publizierens in den Quantenwissenschaften},
   author={Jandura, Sven and Pupillo, Guido},
   year={2022},
   month=may, pages={712} }

@article{Noh_2021,
   title={Operator growth in the transverse-field Ising spin chain with integrability-breaking longitudinal field},
   volume={104},
   ISSN={2470-0053},
   url={http://dx.doi.org/10.1103/PhysRevE.104.034112},
   DOI={10.1103/physreve.104.034112},
   number={3},
   journal={Physical Review E},
   publisher={American Physical Society (APS)},
   author={Noh, Jae Dong},
   year={2021},
   month=sep }

@article{Zhang_2020,
   title={Submicrosecond entangling gate between trapped ions via Rydberg interaction},
   volume={580},
   ISSN={1476-4687},
   url={http://dx.doi.org/10.1038/s41586-020-2152-9},
   DOI={10.1038/s41586-020-2152-9},
   number={7803},
   journal={Nature},
   publisher={Springer Science and Business Media LLC},
   author={Zhang, Chi and Pokorny, Fabian and Li, Weibin and Higgins, Gerard and Pöschl, Andreas and Lesanovsky, Igor and Hennrich, Markus},
   year={2020},
   month=apr, pages={345–349} }

@article{de_Oliveira_2025,
   title={Demonstration of Weighted-Graph Optimization on a Rydberg-Atom Array Using Local Light Shifts},
   volume={6},
   ISSN={2691-3399},
   url={http://dx.doi.org/10.1103/PRXQuantum.6.010301},
   DOI={10.1103/prxquantum.6.010301},
   number={1},
   journal={PRX Quantum},
   publisher={American Physical Society (APS)},
   author={de Oliveira, A. G. and Diamond-Hitchcock, E. and Walker, D. M. and Wells-Pestell, M. T. and Pelegrí, G. and Picken, C. J. and Malcolm, G. P. A. and Daley, A. J. and Bass, J. and Pritchard, J. D.},
   year={2025},
   month=jan }

@article{Semeghini_2021,
   title={Probing topological spin liquids on a programmable quantum simulator},
   volume={374},
   ISSN={1095-9203},
   url={http://dx.doi.org/10.1126/science.abi8794},
   DOI={10.1126/science.abi8794},
   number={6572},
   journal={Science},
   publisher={American Association for the Advancement of Science (AAAS)},
   author={Semeghini, G. and Levine, H. and Keesling, A. and Ebadi, S. and Wang, T. T. and Bluvstein, D. and Verresen, R. and Pichler, H. and Kalinowski, M. and Samajdar, R. and Omran, A. and Sachdev, S. and Vishwanath, A. and Greiner, M. and Vuletić, V. and Lukin, M. D.},
   year={2021},
   month=dec, pages={1242–1247} }

@article{Kornja_a_2023,
   title={Trimer quantum spin liquid in a honeycomb array of Rydberg atoms},
   volume={6},
   ISSN={2399-3650},
   url={http://dx.doi.org/10.1038/s42005-023-01470-z},
   DOI={10.1038/s42005-023-01470-z},
   number={1},
   journal={Communications Physics},
   publisher={Springer Science and Business Media LLC},
   author={Kornjača, Milan and Samajdar, Rhine and Macrì, Tommaso and Gemelke, Nathan and Wang, Sheng-Tao and Liu, Fangli},
   year={2023},
   month=dec }

@article{Giudici_2023,
  title = {Dynamical Preparation of Quantum Spin Liquids in Rydberg Atom Arrays},
  author = {Giudici, Giuliano and Lukin, Mikhail D. and Pichler, Hannes},
  journal = {Phys. Rev. Lett.},
  volume = {129},
  issue = {9},
  pages = {090401},
  numpages = {5},
  year = {2022},
  month = {Aug},
  publisher = {American Physical Society},
  doi = {10.1103/PhysRevLett.129.090401},
  url = {https://link.aps.org/doi/10.1103/PhysRevLett.129.090401}
}

@article{Nguyen_2023,
   title={Quantum Optimization with Arbitrary Connectivity Using Rydberg Atom Arrays},
   volume={4},
   ISSN={2691-3399},
   url={http://dx.doi.org/10.1103/PRXQuantum.4.010316},
   DOI={10.1103/prxquantum.4.010316},
   number={1},
   journal={PRX Quantum},
   publisher={American Physical Society (APS)},
   author={Nguyen, Minh-Thi and Liu, Jin-Guo and Wurtz, Jonathan and Lukin, Mikhail D. and Wang, Sheng-Tao and Pichler, Hannes},
   year={2023},
   month=feb }

@misc{scrambling_2024,
      title={Observation of anomalous information scrambling in a Rydberg atom array}, 
      author={Xinhui Liang and Zongpei Yue and Yu-Xin Chao and Zhen-Xing Hua and Yige Lin and Meng Khoon Tey and Li You},
      year={2024},
      eprint={2410.16174},
      archivePrefix={arXiv},
      primaryClass={quant-ph},
      url={https://arxiv.org/abs/2410.16174}, 
}

@article{Hashizume_2021,
   title={Deterministic Fast Scrambling with Neutral Atom Arrays},
   volume={126},
   ISSN={1079-7114},
   url={http://dx.doi.org/10.1103/PhysRevLett.126.200603},
   DOI={10.1103/physrevlett.126.200603},
   number={20},
   journal={Physical Review Letters},
   publisher={American Physical Society (APS)},
   author={Hashizume, Tomohiro and Bentsen, Gregory S. and Weber, Sebastian and Daley, Andrew J.},
   year={2021},
   month=may }

@article{Serret_2020,
   title={Solving optimization problems with Rydberg analog quantum computers: Realistic requirements for quantum advantage using noisy simulation and classical benchmarks},
   volume={102},
   ISSN={2469-9934},
   url={http://dx.doi.org/10.1103/PhysRevA.102.052617},
   DOI={10.1103/physreva.102.052617},
   number={5},
   journal={Physical Review A},
   publisher={American Physical Society (APS)},
   author={Serret, Michel Fabrice and Marchand, Bertrand and Ayral, Thomas},
   year={2020},
   month=nov }

@article{de_Leseleuc_2019,
   title={Observation of a symmetry-protected topological phase of interacting bosons with Rydberg atoms},
   volume={365},
   ISSN={1095-9203},
   url={http://dx.doi.org/10.1126/science.aav9105},
   DOI={10.1126/science.aav9105},
   number={6455},
   journal={Science},
   publisher={American Association for the Advancement of Science (AAAS)},
   author={de Léséleuc, Sylvain and Lienhard, Vincent and Scholl, Pascal and Barredo, Daniel and Weber, Sebastian and Lang, Nicolai and Büchler, Hans Peter and Lahaye, Thierry and Browaeys, Antoine},
   year={2019},
   month=aug, pages={775–780} }

@article{Bluvstein_2023,
   title={Logical quantum processor based on reconfigurable atom arrays},
   volume={626},
   ISSN={1476-4687},
   url={http://dx.doi.org/10.1038/s41586-023-06927-3},
   DOI={10.1038/s41586-023-06927-3},
   number={7997},
   journal={Nature},
   publisher={Springer Science and Business Media LLC},
   author={Bluvstein, Dolev and Evered, Simon J. and Geim, Alexandra A. and Li, Sophie H. and Zhou, Hengyun and Manovitz, Tom and Ebadi, Sepehr and Cain, Madelyn and Kalinowski, Marcin and Hangleiter, Dominik and Bonilla Ataides, J. Pablo and Maskara, Nishad and Cong, Iris and Gao, Xun and Sales Rodriguez, Pedro and Karolyshyn, Thomas and Semeghini, Giulia and Gullans, Michael J. and Greiner, Markus and Vuletić, Vladan and Lukin, Mikhail D.},
   year={2023},
   month=dec, pages={58–65} }

@article{Schachenmayer_2015,
   title={Thermalization of strongly interacting bosons after spontaneous emissions in optical lattices},
   volume={2},
   ISSN={2196-0763},
   url={http://dx.doi.org/10.1140/epjqt15},
   DOI={10.1140/epjqt15},
   number={1},
   journal={EPJ Quantum Technology},
   publisher={Springer Science and Business Media LLC},
   author={Schachenmayer, Johannes and Pollet, Lode and Troyer, Matthias and Daley, Andrew J},
   year={2015},
   month=jan }

@article{Zeybek_2023,
  title = {Quantum Phases from Competing Van der Waals and Dipole-Dipole Interactions of Rydberg Atoms},
  author = {Zeybek, Zeki and Mukherjee, Rick and Schmelcher, Peter},
  journal = {Phys. Rev. Lett.},
  volume = {131},
  issue = {20},
  pages = {203003},
  numpages = {6},
  year = {2023},
  month = {Nov},
  publisher = {American Physical Society},
  doi = {10.1103/PhysRevLett.131.203003},
  url = {https://link.aps.org/doi/10.1103/PhysRevLett.131.203003}
}

@online{pureDOI_TK,
title={Data for "Simulations of adiabatic state preparation experiments in Rydberg quantum simulators with added realistic laser phase noise"},
author = {T. Kozlej},
year = {2025},
howpublished = {Published using the University of Strathclyde Pure Portal},
doi = {https://doi.org/10.15129/ebc3debb-4360-4195-9908-e7edc092ba97}
}

\appendix

\section{Noise Generation \label{NoiseGEN}}

The method used for noise generation is based on the TK95 algorithm \cite{TK95} for the generation of independent time signals from a frequency power spectrum by applying a complex Gaussian filter followed by the inverse Fourier transform. To begin with, a sample spectral density of frequency noise fluctuations $S_\nu$ is chosen (see Figure \ref{fig:My PSD}).d The frequency spectral density can be transformed into a phase spectral density by applying $S_\nu=\nu^2 S_\phi$. In case a voltage power spectral density is provided $S_V$, we refer the reader to \cite{Schmid_2019} where the conversion to $S_\phi$ is discussed in detail. To create a real noise signal, a double-sided conjugate symmetric spectrum must be generated. Given that the provided spectrum includes only positive frequencies, it is necessary to sample the spectrum twice using a complex Gaussian factor that is conjugated for negative frequencies. A frequency spectrum $S_n$ sampled from the phase noise power spectrum $S_\phi$ is defined by

\begin{align}
S_{n}\bigg[\frac{\mathrm{rad}^{2}}{\mathrm{MHz}}\bigg]^{\frac{1}{2}} & =\mathcal{N}(0)\sqrt{\frac{S_{\phi}}{2}}+i\mathcal{N}(0)\sqrt{\frac{S_{\phi}}{2}},\\
S_{n}(-\nu) & =S_{n}^{*}(\nu)\qquad k\in[0,\mathrm{\nu}_{Nyq}]. \nonumber
\end{align}

\noindent where the relevant units are shown in square brackets and $\mathcal{N}(0)$ is a Gaussian random number used to sample $S_\phi$ at each frequency $\pm k$ up until the Nyquist frequency $\nu_{Nyq}$. Once a trajectory of a phase noise spectrum $S_n$ is generated, the corresponding phase noise signal $\phi(t)$ is derived by an inverse Fourier transform 

\begin{equation}
    \phi(t)=N\frac{\sqrt{2 \Delta \nu }}{2\pi}\sum^{N-1}_{\nu=0} S_n(\nu) e^{2\pi i \nu t /N}.
\end{equation}

\noindent with the addition of a normalization factor where $N$ is the total number of frequency bins $k$ in $S_{w}$ (positive and negative) and a factor of $(2\pi)^{-1}$ is added to account for the Fourier transform in frequency space as opposed to angular frequency. The result of this algorithm is thus a time signal of phase noise $\phi(t)$ in angular units that can then be applied directly to the time evolution of the many-body Hamiltonian in Eq. \eqref{Ryd Hamil}. To test the validity of the generated noise, we compare the autocorrelation function taken directly from the sampled spectrum $A_{S_\phi}(\tau)$ with the averaged autocorrelation across 100 generated noise signals $A_{\phi}$. For this we use the fact that the autocorrelation function can be attained via direct Fourier transform of a given spectral density. Given that $S_{\phi}$ is provided as a discrete series in frequency space, approximating the inverse Fourier transform to a Riemann sum gives

\begin{figure}[t]
\includegraphics[scale=0.4]{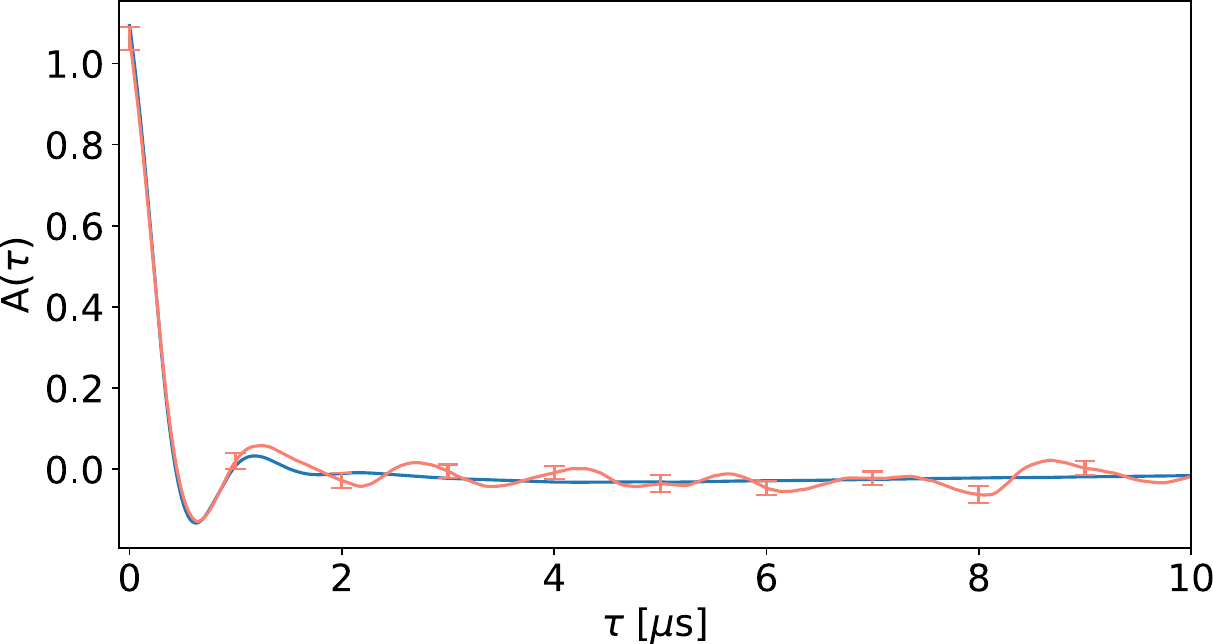}
 \caption{\label{Autocorr} Comparison of the autocorrelation function of the original power density spectrum $A_{S_\phi}$ (blue line) and an average of 100 autocorrelations corresponding to independently  generated noise signals $A_\phi $ (pink line). \justifying  Error bars represent standard error and show good agreement in average noise behavior.}
\end{figure}

\begin{align}
   A_{S_\phi}(\tau)&=\frac{N\Delta \nu}{2\pi}\sum_{\nu} S_\phi(\nu)e^{2\pi i \nu \tau/N}
,\end{align}

\noindent where $A_{S_{\phi}}(\tau)$ is the autocorrelation function of the noise at a given time lag $\tau$. In the case of atocorrelations for the generated phase noise, the autocorrelation of a given time series can be calculated using

\begin{align}
A_{\phi}(\tau)= & \langle\phi_{0}\phi_{\tau}\rangle=\frac{E[(\phi_{0}-E[\phi_{0}])(\phi_{\tau}-E[\phi_{\tau}])]}{\sigma(\phi_{\tau})\sigma(\phi_{0})},
\end{align}

\noindent where $\phi_0$ and $\phi_\tau$ are the same noise signal starting from $t=0$ and a lag of $t=\tau$ respectively, and $\sigma$ gives the standard deviation. Note that the numerator here is just the covariance between $\phi_0$ and $\phi_\tau$ where $E$ signifies the expected value of the given time series. Figure \ref{Autocorr} shows the agreement in autocorrelation functions of the original power spectrum $S_\phi$ and an average of 100 independently generated noise realizations.

\section{\label{app: Rotating frame}Rotating frame transformations in a Rydberg Hamiltonian with laser phase noise}

The standard representation of laser phase noise in a Rydberg Hamiltonian involves a stochastic phase term coupled to the transverse drive. It is often desirable to transform the system into a frame where the Hamiltonian remains purely real. We begin with the Hamiltonian for a Rydberg atom array in the fixed-frequency rotating frame
\begin{equation}
\label{eq:app_ham_initial}
\hat{H}_{R}(t) = \frac{\Omega(t)}{2}\sum^N_k \left[ e^{-i\phi(t)}\hat{\sigma}_k^+ + e^{i\phi(t)}\hat{\sigma}_k^- \right] - \delta(t) \sum^N_k \hat{n}_k + \sum_{k<l} V_{kl} \hat{n}_k \hat{n}_l,
\end{equation}
where $\hat{\sigma}_k^+ = |1\rangle_k \langle 0|_k$ and $\hat{\sigma}_k^- = |0\rangle_k \langle 1|_k$ are the raising and lowering operators, $\hat{n}_k$ is the number operator counting Rydberg occupation, and $\phi(t)$ is a stochastic variable representing phase fluctuations. To eliminate the phase factor from the Rabi term, we define a time-dependent unitary transformation
\begin{equation}
\hat{U}(t) = \exp\left( -i \phi(t) \sum^N_k \hat{n}_k \right),
\end{equation}
which effectively rotates the basis states to a frame that is in line with the phase jitter of the laser. Under a time-dependent unitary transformation, the effective Hamiltonian in the new frame, $\hat{H}'$, is given by the relation
\begin{equation}
\label{eq:app_trans_law}
\hat{H}' = \hat{U}\hat{H}_{R}\hat{U}^\dagger + i \left( \frac{\partial \hat{U}}{\partial t} \right) \hat{U}^\dagger.
\end{equation}

Applying the first term in this transformation, we observe that the detuning $\delta(t)$ and the dipole-dipole interaction $V_{kl}$ commute with $\hat{U}$ because they are built using only $\hat{n}_k$. Consequently, these terms remain invariant in the new frame. However, the transverse coupling term is modified by the transformation of the raising and lowering operators, where $\hat{U} \hat{\sigma}_k^+ \hat{U}^\dagger = \hat{\sigma}_k^+ e^{i\phi(t)}$ and $\hat{U} \hat{\sigma}_k^- \hat{U}^\dagger = \hat{\sigma}_k^- e^{-i\phi(t)}$. Substituting these into the Rabi term of Eq.~\eqref{eq:app_ham_initial} yields
\begin{equation}
\frac{\Omega(t)}{2} \sum_k \left( e^{-i\phi(t)} e^{i\phi(t)} \hat{\sigma}_k^+ + e^{i\phi(t)} e^{-i\phi(t)} \hat{\sigma}_k^- \right) = \frac{\Omega(t)}{2} \sum^N_k \hat{\sigma}_x^k.
\end{equation}
This step effectively ``unwinds'' the phase noise from the transverse drive. The second part of the transformation law in Eq.~\eqref{eq:app_trans_law} accounts for the ``inertial'' or ``gauge'' term arising from the explicit time-dependence of the unitary operator. Differentiating $\hat{U}$ with respect to time gives
\begin{equation}
\frac{\partial \hat{U}}{\partial t} = -i \dot{\phi}(t) \left( \sum^N_k \hat{n}_k \right) \exp\left( -i \phi(t) \sum^N_k \hat{n}_k \right).
\end{equation}

Multiplying by $i$ and $\hat{U}^\dagger$ results in the addition of a longitudinal term $\dot{\phi}(t) \sum \hat{n}_k$ to the Hamiltonian. Physically, $\dot{\phi}(t)$ represents the instantaneous frequency fluctuation of the laser, and can thus be understood as a stochastic jitter in phase. Combining all components, the final transformed Hamiltonian is expressed as
\begin{equation}
\hat{H}' = \frac{\Omega(t)}{2} \sum^N_k \hat{\sigma}_x^k - \left[ \delta(t) - \dot{\phi}(t) \right] \sum^N_k \hat{n}_k + \sum_{k<l} V_{kl} \hat{n}_k \hat{n}_l.
\end{equation}
This result provides a rigorous mathematical justification for treating laser phase noise as a stochastic fluctuation of the detuning. In the context of numerical modeling using the TK95 algorithm, this transformation allows the use of purely real-valued matrices.

\section{Excitation dynamics \label{noiseEXC}}

Having discussed phase noise generation, we now demonstrate some of the features in the excitation dynamics in our system. Consider again the Hamiltonian in Eq. \ref{Ryd Hamil} where a given noise array $\phi(t)$ appears as a small phase modulation on the Rabi frequency of the laser drive at every time step in the evolution. The following section will first highlight the reflection symmetry that governs the dynamics in this Hamiltonian, as well as exploring the different features in the diagonal ensemble after phase noise excitation in a time independent Hamiltonian.

\subsection{On integrability \label{app: Integrability}}

One of the goals of this analysis will be to reveal how integrability and spectral structure influence the ability of phase noise to spread energy throughout the system. However, before beginning the discussion on excitation dynamics is useful to define what is meant by integrability in the context of the Rydberg chain Ising Hamiltonian we are simulating in this paper, as this term may often be used to cover a wide range of phenomena across different systems. In this work, integrability refers to the regime where the system's dynamics are governed by a complete set of conserved quantities, resulting in a highly structured and `predictable' energy spectrum. In the limit of $\Omega \to 0$, the Hamiltonian reduces to a Ising model where the number of Rydberg excitations is conserved, as without the transverse field there is no mechanism to flip the states of individual sites. Consequently, eigenstates are simply computational basis states, and the ground state is determined by the competition between $\delta$ and $V_{dd}$. In this integrable regime, the system is `stuck' in a specific energy sector with different degenerate configurations with the same number of Rydberg excitations. As the transverse field $\Omega$ is increased, these conservation laws are broken, and the system transitions into a non-integrable, or chaotic, regime. Although the transition to a non-integrable regime is gradual and impossible to pinpoint precisely, it is marked by a break-down of degeneracy as $\Omega$ introduces state mixing and energy state repulsion as the degenerate computational states form superposition states with varying energies. In larger $\Omega$ the once isolated energy sectors merge into one sector with a high density of eigenstates that have distinct energies, making dynamics in such a system much more complex, as different couplings between states across the entire energy spectrum become accessible. Dynamics in such non-integrable systems becomes much more difficult to calculate especially at larger system sizes. This breakdown of integrability acts as a bridge, allowing phase noise to couple the ground state to a dense manifold of excited states that would otherwise be inaccessible in the highly structured and disconnected energy spectrum of the strictly integrable Hamiltonian.

\begin{figure}[t]
\includegraphics[scale=0.4]{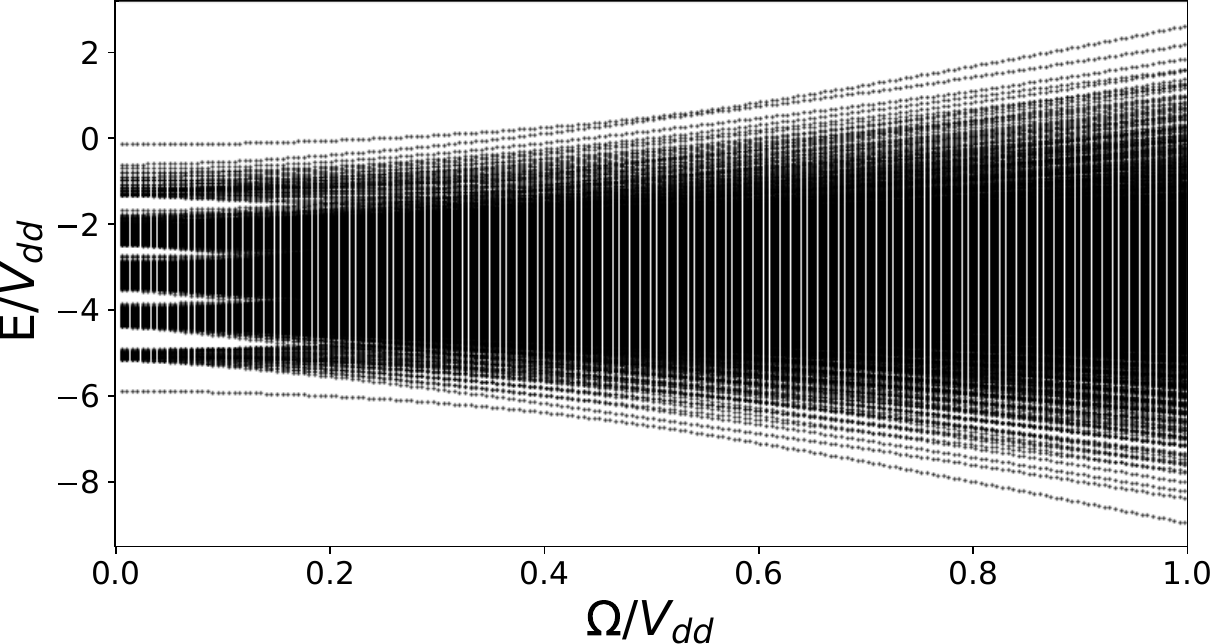}
 \caption{\label{fig: SYMplot} A plot of symmetric energy eigenvalues of the Rydberg Hamiltonian defined in Eq. \eqref{Ryd Hamil} as a function of  Rabi frequency $\Omega$ for a constant detuning of $\delta=1.1V_{dd}$.\justifying  Increasing $\Omega$ strengthens the transverse field, leading to a rapid breakdown in integrability as initially clustered eigenstates spread out and merge.}
\end{figure}

Figure \ref{fig: SYMplot} shows the separation of eigenenergies corresponding to eigenstates as a function of Rabi frequency $\Omega/V_{dd}$ and with constant detuning $\delta=1.1V_{dd}$, which are relevant parameters for the third step in the simulated adiabatic state preparation. Tracing the bottom lone which represents the ground state we see the minimum ground state energy gap corresponding to the critical region of the preparation that occurs between $\Omega=0.2-0.3 V_{dd}$. Furthermore, the total energy span of the Hilbert space increases from $10V_{dd}$ to $6V_{dd}$ as $\Omega/V_{dd}$ increases. Initially, clusters of energy eigenvalues form an underlying structure of clustered energy eigenstates that is consistent with what one might expect in an Ising Hamiltonian, but as $\Omega/V_{dd}$ we see this structure break down as the clusters merge. This merging corresponds to the breaking of integrability that occurs when a transverse field is applied to the Ising Hamiltonian, which has direct implications for phase noise excitation.

\subsection{\label{app:Reflection}Reflection symmetry}

Importantly, this Hamiltonian has a reflection symmetry in the computational basis. Given that we limit ourselves to odd numbers of sites, a reflection operator that flips all sites about the center of the chain can be defined in terms of $\sigma_+=(\sigma_x+i\sigma_y)/2$ and $\sigma_-=(\sigma_x-i\sigma_y)/2$ operators as

\begin{flalign}
\label{eq: Reflector}
\hat{R} = \prod_{i=1}^{ N/2 } \bigg( &\hat{\sigma}_+^{(i)} \hat{\sigma}_-^{(N-i+1)} + \hat{\sigma}_-^{(i)} \hat{\sigma}_+^{(N-i+1)} \\&+ \frac{1}{2} \big(1 + \hat{\sigma}_z^{(i)} \hat{\sigma}_z^{(N-i+1)}\big)\bigg) .\nonumber
\end{flalign}

\noindent The first two terms in the operator flip sites at either end of the chain while the third term ensures the preservation of the spin alignment if both sites are in the same state. The Hamiltonian can be shown to commute with this reflection operator, leading to two independent symmetry sectors of eigenstates made up of symmetric and anti-symmetric superpositions of computational states. For open boundary one-dimensional spin chains with an odd number of sites, there will always be computational states that are invariant under reflection. For example, the crystalline states discussed in Figure \ref{fig:Ryd Adiab Sketch} that dominate the ground state for low values of $\Omega$. Such reflection invariant computational states are only represented in the symmetric energy eigenstates. Furthermore, it can be shown that diabatic excitation, governed by changes to $\Omega$, as well as phase noise excitation, governed by changes in $\phi$, both respect this symmetry. This means that if all energy is initially in one of these symmetry sectors as is the case in the symmetric ground state at the beginning of the adiabatic state preparation, all dynamics will be confined to that symmetry sector. This effectively inhibits excitation in approximately half of the Hilbert space, thus halving the density of states that the noise can dissipate into. Consequently, Figure \ref{fig: SYMplot} only shows symmetric eigenstates relevant to our simulations.

\subsection{\label{AppB: TI phase noise}Laser phase noise excitations in time independent Hamiltonians}

\begin{figure*}
  \includegraphics[scale=0.4]{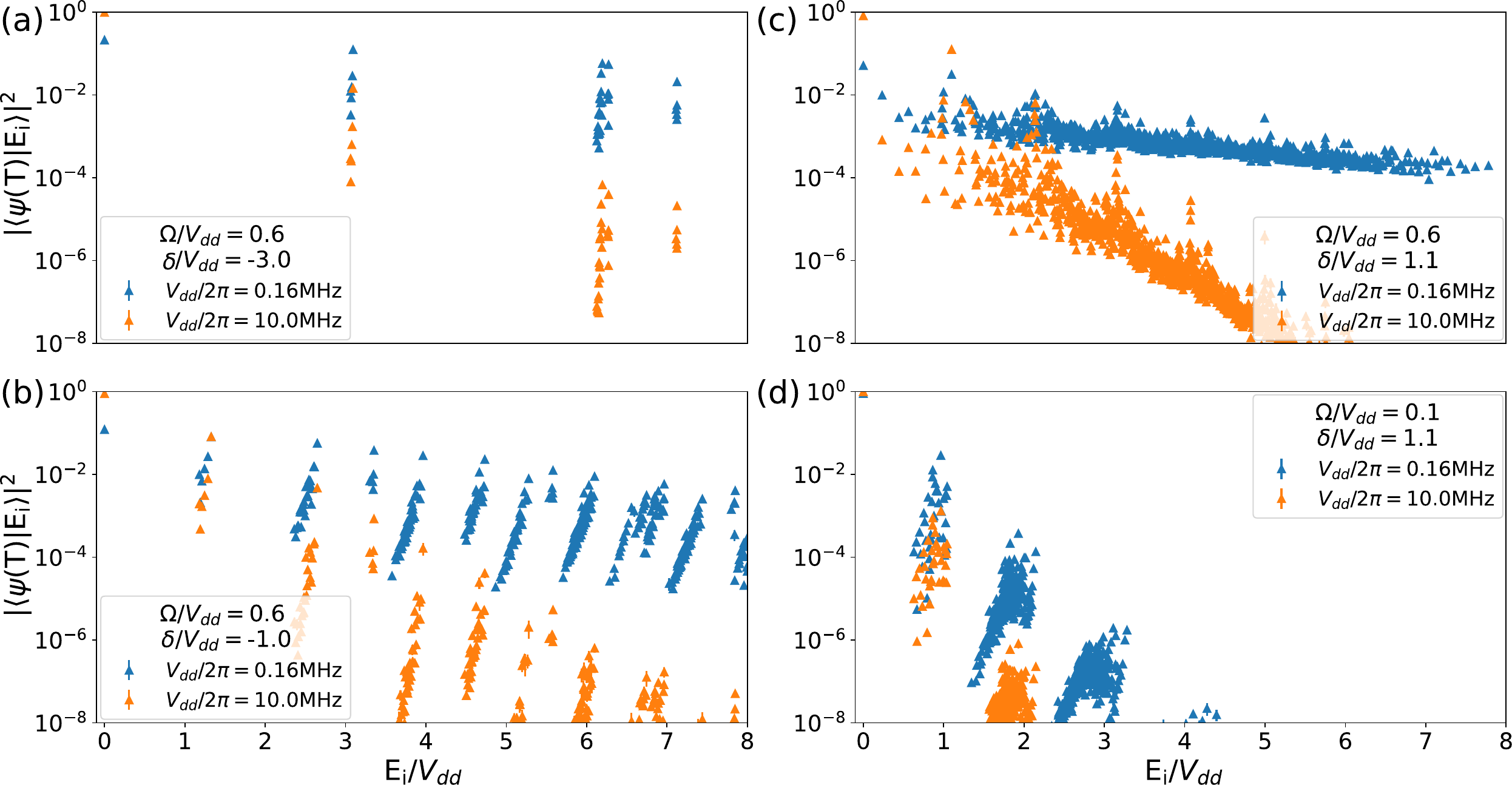}
  \caption{ \label{fig: pDE const EVO kappaCMP}  The ground state of the Hamiltonian from Eq. \eqref{Ryd Hamil} is evolved with noise for a total time of $TV_{dd}=300$, for different laser parameters  (a) $\Omega/V_{dd}=0.6$ and $\delta/V_{dd}=-3.0$, (b) $\Omega/V_{dd}=0.6$ and $\delta/V_{dd}=-1.0$, (c) $\Omega/V_{dd}=0.6$ and $\delta/V_{dd}=1.1$, (d) $\Omega/V_{dd}=0.1$ and $\delta/V_{dd}=1.1$. \justifying In orange we have the diagonal ensemble after noisy evolution with site interaction strength set to $V_{dd}/2\pi=10$MHz, while in blue we have the same simulation reproduced for an interaction strength of $V_{dd}/2\pi=0.16$MHz. Error bars plot standard error across 100 time evolutions with independent laser phase noise signals using the same noise profile from Figure \ref{fig:My PSD}, but are too small to be seen on the log scale suggesting a consistent excitation pattern.}
\end{figure*}

While it is simple to treat the adiabatic state preparation as a black box and analyze only the final state, getting a deeper understanding of the complicated excitation dynamics of laser phase noise a time dependent Hamiltonian with ever changing energy level structure is much harder. The transition rates discussed
in Figure \ref{fig: Matrix Elements} provide a road map for a given source of energy to excite the system
to higher many-body states. In order to bring these transition rate snapshots to
life, in this section we perform an analysis of excitation dynamics using the same time-independent Hamiltonians.

Figure \ref{fig: pDE const EVO kappaCMP} shows the diagonal ensembles $|\langle \psi(T_3)|E_i \rangle|^2$ for state $|\psi(T_3)\rangle$ after a ground state evolution of the time-independent Hamiltonians with energy eigenstates $|E_i\rangle$ and laser parameters that occur at different stages of the adiabatic state preparation analyzed in Section \ref{Sec 3}, with the corresponding matrix element analysis in Figure \ref{fig: Matrix Elements}. Simulations are performed for two separate interaction strength regimes $V_{dd}/2\pi=10,0.159$ MHz, which are studied in Sections \ref{Sec 3} and \ref{Sec 4} respectively. All simulations are averaged across 100 independent noise realizations generated from the same power spectrum in Figure \ref{fig:My PSD} as described in Appendix \ref{NoiseGEN}. To ensure both interaction regimes have equivalent dynamics when no phase noise is added, simulations are performed for a fixed total evolution time of $TV_{dd}=300$, which corresponds to completing the protocol in $\approx4.8\mu$s for the fast $10$MHz regime and $300\mu$s for the slow $0.159$MHz regime.

Figure \ref{fig: pDE const EVO kappaCMP}(a) shows the noisy evolution of the ground state of a Hamiltonian that occurs at the beginning of stage 2 of the adiabatic protocol. The resulting excitation is very low in the $V_{dd}/2\pi=10$MHz regime with a final ground state occupation of 98.2\% at a standard error of 0.2\%, but much larger in the $V_{dd}/2\pi=0.16$MHz regime in which ground state population dropped to 21.1\% with standard error of $3$\%, suggesting that proximity of the dynamical frequency of the system to the $0.48$MHz peak is a key factor for driving excitation even in the far detuned regime with a low density of states with high transition rates. Figure \ref{fig: pDE const EVO kappaCMP}(b) with laser parameters from the middle of stage 2 paints a similar picture but with a modest increase in excitation across the spectrum as the energy gap between states decreases and more states become viable for excitation. The $V_{dd}/2\pi=10$MHz regime here has a final ground state population $89.4$\% at a standard error of 2\%, while the ground state of the $V_{dd}/2\pi=0.16$MHz regime drops further to $12.1$\% with a standard error increasing to $4$\%. In particular, energy in the $V_{dd}/2\pi=0.16$MHz regime is much more evenly distributed throughout the energy spectrum than in the case of the $V_{dd}/2\pi=10$MHz regime, which has a much higher drop-off in occupation of higher energy states as most energy is still concentrated in the ground state and states with the highest ground state transition rate. As predicted in the matrix element analysis, Figure \ref{fig: pDE const EVO kappaCMP}(c) that uses a parameters from the transition point between stages 2 and 3 results in the most overall excitation, with a ground state occupancy of $81.5$\% at a $3$\% standard error for the $V_{dd}/2\pi=10$MHz regime, and an almost fully depleted ground state occupancy of $5.1$\% at a $1$\% standard error for the $V_{dd}/2\pi=0.16$MHz regime. The evolution of the $V_{dd}/2\pi=10$MHz regime results in the formation of well-defined equidistant peaks as the energy of the phase noise occupancy cascades from one high transition rate to the next with a gradually decreasing probability. While these peaks are still visible in the $V_{dd}/2\pi=0.16$MHz regime, the comparatively stronger effect of phase noise leads to higher energy transfer with occupancy largely equilibrating across the entire spectrum. All three of these figures with a high transverse field at $\Omega/V_{dd}=0.6$  have excitation profiles dominated by transitions that are delocalized in energy, allowing noise to access the entire energy spectrum over time. Moreover, the rate of energy transfer improves as $\delta$ becomes increasingly positive and the energy required for transitions between many-body states with different numbers of atoms in a Rydberg state drops (see Figure \ref{fig: SYMplot}). However, for the Hamiltonian that occurs at the end of the protocol seen in Figure \ref{fig: pDE const EVO kappaCMP}(d), in which $\delta$ is positive but the transverse field drops down to $\Omega/V_{dd}=0.1$, there is a dramatic decrease in excitation across both regimes as the final ground state occupancy is $99.4$\% and $90.3$\% for the $V_{dd}/2\pi=10$MHz and $V_{dd}/2\pi=0.16$MHz regimes, respectively. The main difference is that the system here is much more integrable, leading to available transition rates that are not only lower but also highly localized in energy, meaning that phase noise transfers energy from the ground state at a much lower rate and is restricted to low energy sectors.

\end{document}